\documentclass[preprint, 3p]{elsarticle}
  
\usepackage{url}
\usepackage[pagewise]{lineno}
\usepackage{graphicx}
\usepackage{amsmath, amssymb}
\usepackage{algorithm, algpseudocode} 
\usepackage[standard]{ntheorem}

\usepackage[utf8]{inputenc} 
\usepackage{pgfplots}
\pgfplotsset{compat=newest}
\usepgfplotslibrary{groupplots}
\usetikzlibrary{matrix,positioning}
\usepgfplotslibrary{dateplot}

\newcommand{\mat}[1]{\mathbf{#1}}
\renewcommand{\vec}[1]{\boldsymbol{#1}}
\renewcommand{\det}[1]{\lvert #1 \rvert}

\newcommand{\Z}{\mat{Z}}
\newcommand{\W}{\mat{W}}

\newcommand{\Esp}[2]{\mathbb{E}_{#1}\left[#2 \right]}
\renewcommand{\v}{\vec{v}}
\newcommand{\eye}[1]{\mat{I}_{#1}}
\newcommand{\norm}[1]{\| #1 \|}

\newcommand{\rev}[1]{\color{black}#1 \color{black}}
 
\def\y{{\vec{y}}}
\def\u{{\vec{u}}}
\newcommand{\Q}[1]{Q_{#1}(\vec{\theta} \mid \vec{\theta}^{(m)})}
\def\uihat{{\hat{\vec{u}}_i }}
\def\yhat{{\hat{\vec{y}} }}

\def\Sigmahat{{\hat{\mat{\Sigma}}}}

\begin{document}

\begin{frontmatter}

  \title{Multifrequency Array Calibration in Presence of Radio Frequency Interferences}

  \author[1]{Yassine Mhiri \corref{cor1}}% \fnref{fn1}}
  \ead{yassine.mhiri@universite-paris-saclay.fr}

  \author[2]{Mohammed Nabil El Korso}% \fnref{fn1}}
  \ead{m.elkorso@parisnanterre.fr}
  \author[2]{Arnaud Breloy}
  \ead{a.breloy@parisnanterre.fr}
  \author[1]{Pascal Larzabal}
  \ead{pascal.larzabal@universite-paris-saclay.fr}

  \cortext[cor1]{Corresponding author}
  %\fntext[fn1]{is an EURASIP member}
  % footnote and it should be a really long footnote. But this
  % footnote is not yet sufficiently long enough to make two
  % lines of footnote text.}
  % \fntext[fn3]{Yet another author footnote.}
  % \fntext[fn4]{Yet another author footnote.}

  \address[1]{SATIE, CNRS, ENS Paris-Saclay, Université Paris-Saclay
  ,4 Avenue des Sciences 91190 Gif-sur-Yvette,
  France}
  \address[2]{LEME, Université Paris Nanterre, 50 rue de Sèvres
  92410 Ville d’Avray, France}

  %\markboth{Journal of \LaTeX\ Class Files, Vol. 14, No. 8, August 2015}
  %{Shell \MakeLowercase{\textit{et al.}}: Bare Demo of IEEEtran.cls for IEEE Journals}
  \begin{abstract}

  Radio interferometers are phased arrays producing high-resolution images from the covariance matrix of measurements.
  Calibration of such instruments is necessary and is a critical task. This is how the estimation of instrumental errors is 
  usually done thanks to the knowledge of referenced celestial sources. However, the use of high sensitive antennas in modern 
  radio interferometers (LOFAR, SKA) brings a new challenge in radio astronomy because they are more sensitive to Radio Frequency Interferences (RFI). 
  The presence of RFI during the calibration process generally induces biases in state-of-the-art solutions.
  The purpose of this paper is to propose an alternative to alleviate the effects of RFI. 
  For that, we first propose a model to take into account the presence of RFI in the data across multiple
  frequency channels thanks to a low-rank structured noise. We then achieve maximum likelihood estimation of
  the calibration parameters with a Space Alternating Generalized Expectation-Maximization (SAGE) algorithm for which we derive originally two sets of complete data allowing 
  \rev{closed-form} expressions for the updates. Numerical simulations show a significant gain in performance for RFI corrupted 
  data in comparison with some more classical methods.

  \end{abstract}

  \begin{keyword}
  Radio astronomy\sep calibration\sep SAGE\sep RFI mitigation\sep array signal processing 
  \end{keyword}

\end{frontmatter}
\section{Introduction}

\rev{Radio astronomy} uses radio telescopes to detect weak emissions from celestial sources in the radio spectrum.  
Since the first radio observation of the sky in 1931 by Karl Jansky \cite{Jansky1933}, we have been able to monitor the sky in unprecedented detail
with growing resolution, bringing new insights in various scientific domains such as solar monitoring, planetology, and astrophysics.
Moreover, there has been major improvements in the technology used by radio interferometers, making possible the first imaging of a black hole \cite{M87}.
A radio interferometer is an array of antennas that are placed far apart from each other.  Thanks to the correlations (called visibilities) between signals 
at each sensor it produces a high-resolution image of the sky \cite{Thompson17}.
A new generation of radio interferometers, such as the LOFAR or the SKA, are designed to further improve the sensitivity of 
radio observations over a large bandwidth using multiple sensors placed on a large area.
\rev{These technological advances bring} several array processing challenges to exploit the full scientific
potential of such instruments. In this paper, we consider the problem of array calibration, which is a necessary step to produce reliable observations of the sky with high sensitivity.

Array calibration consists in finding an estimate of the systematic instrumental errors affecting the measurements.
Such perturbations range from antenna electronic gains to atmospheric perturbations and 
can be modeled using the Jones framework \cite{Smirnov11}\cite{HBS1996}. Array calibration methods can usually be classified into two types of approaches depending on the 
presence or not of calibrator sources. Radio interferometer calibration lies in the first category where we rely on the position and flux of strong celestial sources as calibrators to estimate 
the parameters of the model.
The field of radio astronomy has seen several successive generations of calibration algorithms.  %1GC, 2GC, 3GC, 4GC
The latest tendency brings the focus on statistical modeling by the formulation of robust calibration algorithms as the
solution of a maximum likelihood problem. 
In the literature, we can find numerous non-linear least square approaches \cite{BV03}\cite{WV09}\cite{SW14}\cite{tasse14} that assume an additive 
Gaussian noise on the visibilities. Some alternate modeling of the noise has been proposed, notably, a Student's $t$-distribution \cite{Kazemi13}\cite{Ollier17}\cite{SBSKG20}\rev{\cite{Mhiri2021expectation}\cite{yatawatta2020stochastic}}  to take into account the eventual
presence of outliers. 

One of the main limitations to the sensitivity of radio telescopes is Radio Frequency Interferences (RFI) which are mostly due to
man-made radio waves. RFI can come from various sources, from telecommunication signals
to high voltage transmission lines. Moreover, the generalization of large scale telecommunication networks and the growing use of satellites makes
the modeling and mitigation of RFI a significant challenge in array signal processing.
As a matter of fact, despite the efforts that are made to build such instruments in radio-quiet zones (in which radio transmissions are rare or restricted),
man-made radio waves still have a significant 
impact in radio astronomy observation \cite{Baan19}. Likewise, the new generation of large-scale radio interferometers
can detect weaker signals thanks to sensors being more and more sensitive, increasing the chances of intercepting RFI signals.
The presence of a strong RFI in a radio interferometer signal corrupts the data measured by the affected sensors. Such outliers are usually flagged in the raw data using a 
generalized likelihood ratio test \cite{Offringa10} or by a manual inspection of the data to check for its consistency. 
Various methods to perform spatial filtering on the correlation matrix have been proposed \cite{LV00}. They are particularly efficient when the power of the RFI is sufficiently
large so that the subspace of the interfering signal is identifiable. However, they lose in performance when the power of the RFI decreases.
Yet, the presence of RFI still significantly impact the calibration solutions making the data 
unusable for scientific purposes \cite{boonstra05}.

In this paper, we propose a unified model from which we derive a maximum likelihood estimate of the calibration solutions as well as the RFI subspace, 
allowing robust calibration in the presence of RFI.
By modeling accurately the RFI contribution to the visibilities, we improve the robustness
of the calibration process for weak and medium RFI. 
Considering RFI as man-made high-frequency electromagnetic waves \cite{Lemmon97}\cite{LV00}, we prove that the subspace of the perturbing RFI 
yields a low-rank structured perturbation in the visibilities. Consequently, we build a model for multi-frequency observations
using the formalism of mixed effect model \cite{lavielle14}. We consider a model composed of
fixed effects representing the observed celestial sources and random effects which account for the possible presence of RFI exclusively on a few frequency channels. 
We achieve maximum likelihood estimation for the introduced model by using the Space Alternating Generalized EM (SAGE) \cite{Fessler94}.
The proposed model is described in section 2. In section 3, we introduce a SAGE-based algorithm used to perform maximum likelihood estimation for the proposed model. 
Finally, numerical results are presented in section 4.

% Ajouter une petite phrase sur la modelisation des frèquence  
% Otherwise, classical calibration algorithms do not take into account the diversity of the systematic errors across frequency bands, 
% making the number of unknown parameters grow with the number of frequency channels. A physical model for
% the frequency dependence exists for specific errors \cite{Gasperin19}. For sake of simplicity, it is commonly accepted \cite{Yatawatta15} to enforce the smoothness 
% of the calibration parameters across the frequency channels with a power law, reducing the number of unknown parameters while considering the 
% frequency dependence of the systematic errors. On the other hand, we cannot assume the presence 
% or absence of weak RFI in a specific frequency band, making calibration over a large bandwidth in presence of RFI difficult.

\textit{Notation} : Matrices are represented in bold uppercase and vectors in bold lowercase.
The expectation 
operator is denoted  $\Esp{.}{.}$. The transpose operator is noted $^T$, the complex conjugate $^*$ and the complex conjugate transpose (Hermitian)
$^H$. $\otimes$ denotes the Kronecker operator. The vectorization of a matrix, vec(.), is done by stacking its columns in a vector
and its inverse operator is denoted unvec(.). 
$\Re\{.\}$ and $\Im\{.\}$ respectively refer to the real and imaginary
part of a complex. The \rev{cardinality} of a set is denoted Card(.). diag(.) creates a block-diagonal matrix from its matrix argument, rank(.) computes the rank of a matrix, Tr(.) refers to
the trace of a matrix, and $|.|$ to its determinant. The Frobenius norm is denoted $\norm{.}_2^2$. $\sim$ "means distributed as", $\overset{d}{=}$ stands for "shares the
same distribution" and $\overset{d}{\rightarrow}$ denotes 
convergence in distribution. $\mathcal{CN}(.,.)$ refers to the complex normal distribution,
$\mathcal{GCN}(.,.,.)$ to the generalized complex normal distribution, $\mathcal{W}(.,.)$ to the Wishart distribution and $\mathcal{U}(.,.)$ to the uniform distribution.

\section{Data model} 
We consider a radio interferometer composed of $P$ receivers observing the sky, each receiver being dual-polarized such that 
the signal received by the $p^{th}$ sensor at time $t$ can be expressed as

\begin{equation}
\vec{x}_p(t) = \begin{bmatrix} x_{p,H}(t) \\ x_{p,V}(t) \end{bmatrix}
\end{equation}
where $x_{p,H}(t)$ and $x_{p,V}(t)$ represent the horizontal and vertical polarizations.
\newline   
In the context of radio interferometer calibration, the sky is usually modeled as the sum of $D$ independant strong calibrator sources,
each source decomposed 
into two orthogonal polarizations \cite{WB06}\cite{Kazemi11}\cite{Smirnov11}\cite{SBSKG20}. 
The known sources, $\vec{s}(t)$, are assumed to be sampled from a circular complex gaussian law with a brightness (or coherency) matrix $\mat{C}$, ie,
\begin{equation}
\vec{s}(t) = \begin{bmatrix} s_{H}(t) \\ s_{V}(t) \end{bmatrix} \sim \mathcal{CN}(\vec{0}, \mat{C}), \ \mbox{in which} \ \mat{C} = \begin{bmatrix} I+Q & U+jV \\ U-jV & I-Q\end{bmatrix}, 
\end{equation}
where $I, Q, U, V$ are the Stokes parameters of the source caracterizing its power and polarization \cite{Thompson17}.

The perturbations encountered by the signal from its emission to its reception by the telescope sensors are modeled by 
a matrix product of the emitted signal with a perturbation matrix, namely the Jones matrix. 
Jones matrices can accurately model all kinds of distorsions of the signal (more details on the Jones framework 
can be found in \cite{Smirnov11}\cite{HBS1996}\cite{Thompson17}). A Jones matrix is specified for each receiver $p$ and source $i$  such that the signal received at the $p^{th}$ sensor can be
ideally written as
\begin{equation}
\vec{x}_p(t) = \sum_{i=1}^D \mat{J}_{i,p}(t) \vec{s}_i(t).
\end{equation}

\subsection{Radio interferometric data model}

Radio interferometry uses estimates of the correlations between sensors to recover the flux of celestial sources \cite{Thompson17}\cite{WB06}.
By concatenating the signals from all sensors in a vector 
$\vec{x}(t) = \begin{bmatrix} \vec{x}_1^T(t) & \hdots &\vec{x}_P^T(t) \end{bmatrix}^T$
The correlation matrix of the array can be defined as a $P^2$ $2\times2$ block matrix, $\mat{V} = \Esp{}{\vec{x}(t)\vec{x}^H(t)}$
where each block is the correlation between the $p^{th}$ and the $q^{th}$ receivers,

\begin{equation}\label{eqn:RIME}
  \mat{V}_{pq} = \Esp{}{\vec{x}_p(t) \vec{x}_q^H(t)} = \sum_{i=1}^D \mat{J}_{i,p}\mat{C}_i\mat{J}_{i,q}^H,
\end{equation}
where $\mat{V}_{pq}$ is referred to as the visibility for the baseline formed by the $p^{th}$ and $q^{th}$ sensors and equation (\ref{eqn:RIME}) is referred to as the Radio Interferometric Measurement Equation (RIME) \cite{Smirnov11}.

%The process of calibration aims at finding a good estimate of the Jones matrices using the observations of known sources.

In order to reduce the memory cost and complexity of calibration algorithms we use a vectorized 
version of the correlation matrix that contains only its lower diagonal entries.
Let us note $\mathcal{B} = \{ (p,q)\in[1,P]^2, p<q \}$ the set of unique pairs in the sensor array 
and $N_\mathcal{B} = \mbox{Card}(\mathcal{B}) = \frac{P(P-1)}{2}$. \\
For $(p,q)\in\mathcal{B}$, the visibility vector for the baseline formed by the 
$p^{th}$ and $q^{th}$ sensors is noted,
\begin{equation}
\vec{v}_{pq} = \mbox{vec}(\mat{V}_{pq}) = \sum_{i=1}^D \left( \mat{J}_{i,q}^* \otimes \mat{J}_{i,p}\right)\mbox{vec}(\mat{C}_i),
\end{equation}
where the subscript $pq$ refers to the pair of antennas formed by the $p^{th}$ and the $q^{th}$ antennas.
All the visibility vectors $\{\vec{v}_{pq}\}_{(p,q)\in \mathcal{B}}$ are then concatenated in a vector
\begin{equation}
\vec{v} = \begin{bmatrix} \vec{v}_{12} \\ \vdots\\ \vec{v}_{(P-1)P}\end{bmatrix}.
\end{equation}

In practice the \rev{visibilities} are estimated by the sample covariance matrix of the measurement vectors for a short time interval,

\begin{equation}
\mat{R}_{pq} = \frac{1}{N} \sum_{n=1}^N \vec{x}_p(n)\vec{x}_q^H(n),
\end{equation}
where $N$ is the number of bins chosen such that the sources signals are stationnary and the Jones matrices constant.
For $N$ sufficiently large it is commonly accepted to model $\mbox{vec}(\mat{R}_{pq})$ as the sum of a deterministic signal
and a white gaussian noise \cite{Thompson17}\cite{Yatawatta15}\cite{Yatawatta20}\cite{WB06}\cite{SBSKG20} as

%electronic gains as well as perturbations and biases induced from the computation of those estimates are modeled by a white complex gaussian noise

\begin{equation}\label{eqn:model-baseline}
\mat{r}_{pq}= \mbox{vec}(\mat{R}_{pq}) \overset{d}{=} \vec{v}_{pq} + \vec{n}_{pq}, \quad \vec{n}_{pq} \sim \mathcal{CN}(0, \sigma^2 \mat{I}_4).
\end{equation}
The additive noise $\vec{n}_{pq}$ is often referred to as the thermal noise, incorporating the noise induced by
the antenna and cables that carry the signal to the correlator \cite{Thompson17}. Moreover, $\vec{r}_{pq}$ being a 
realization of the sample covariance matrix associated to $\vec{v}_{pq}$ in an asymptotic regime, it is convenient
to model it with a complex circular gaussian distribution \cite{mahot2013}.

The general problem of radio interferometric calibration aims at estimating the Jones matrices and the noise parameter given a set of
measurements of the \rev{visibilities} from known stellar sources. Taking into account all the baselines, equation (\ref{eqn:model-baseline}) becomes 
\begin{equation}\label{eqn:model-no-outliers}
\vec{r}= \vec{v} + \vec{n}, \quad \mat{n} \sim \mathcal{CN}(0, \sigma^2 \mat{I}_{N_\mathcal{B}}),
\end{equation}
where $\vec{r}= \begin{bmatrix} \vec{r}_{12} \\ \vdots \\ \vec{r}_{(P-1)P} \end{bmatrix}$, 
$\vec{v}= \begin{bmatrix} \vec{v}_{12} \\ \vdots \\ \vec{v}_{(P-1)P} \end{bmatrix}$.

\subsection{Radio Frequency Interferences}

RFI are jamming signals of various forms (telecommunication, high voltage transmission lines) that can be modeled 
as additional independent unwanted sources with their specific Jones matrices \cite{Lemmon97}. 
The model expressed in (\ref{eqn:model-no-outliers}) is not suited for the presence of RFI. In response,
various works have proposed statistical models to take into account the possible presence of RFI \cite{SBSKG20}\cite{Ollier17}.
Spatial filtering and statistical tests have been used to tackle the possible presence of strong RFI \cite{LV00}\cite{Fridman01}\cite{Baan19}\cite{Offringa10}\cite{Yatawatta20}. 
% preciser : "soit pour rejeter la données soit pour eliminer les rfi de la mesure
Such methods become less effective when the power of the RFI decreases. 
As an alternative we propose to perform calibration in presence of RFI, 
taking into account the RFI contribution to the visibilities in the calibration step.
To this end, we propose in this section a statistical model for the measured visibilities corrupted by RFI from which a Maximum Likelihood Estimate (MLE) is derived in section 3.
% In this section we propose a statistical model for the measured visibilities corrupted by weak RFI.

Let us consider a vector of measured visibilities corrupted by $L$ RFI sources, we propose to model the presence of RFI in the measured visibilities
by a stochastic process with a low-rank structure, $\vec{r}^{RFI} = \mat{W}\vec{y}$. The RFI, the celestial sources, and
the Gaussian noise are considered independent so that the presence of RFI is modeled independently and added to equation 
(\ref{eqn:model-no-outliers}). The model on the measured visibilities thus becomes

\begin{equation}\label{eqn:model-rfi-monofreq}
\vec{r} = \vec{v} + \mat{W}\vec{y} + \vec{n}, \ \vec{y} \sim \mathcal{CN}(\mbox{vec}(\mat{I}), \mat{I}), 
\ \vec{n}\sim \mathcal{CN}(\vec{0},\sigma^2\mat{I}), \ \rho \leq 4L^2,
\end{equation}
where $\mat{W}\in\mathbb{C}^{4N_\mathcal{B} \times 4L^2}$, $\rho =\mbox{rank}(\mat{W})$.
Specifically, the rational behind the modelling of the RFI in (\ref{eqn:model-rfi-monofreq}) is motivated by the 
physical modelling of such man-made signal \cite{Lemmon97}.
Indeed the RFI component of the received signal at the $p^{th}$ sensor can be expressed as follows,

\begin{equation}
\vec{s}^{RFI}_l(t) =  \begin{bmatrix} s^{RFI}_{l,H}(t) \\ s^{RFI}_{l,V}(t) \end{bmatrix} \sim \mathcal{CN}(0, \mat{C}^{RFI}_l)
, \quad \vec{x}^{RFI}_p(t) = \sum_{l=1}^L \mat{J}^{RFI}_{l,p}(t) \vec{s}^{RFI}_l(t).
\end{equation}
The form of the correlation matrix of the RFI sources for the P-sensors 
telescope can be expressed using the RIME as defined in the previous section,

\begin{equation}\label{eqn:RIME-RFI}
  \mat{V}^{RFI}_{pq} = \Esp{}{\vec{x}^{RFI}_p(t) \vec{x}^{RFI^{H}}_q(t)} = \sum_{i=1}^L \mat{J}^{RFI}_{i,p}\mat{C}^{RFI}_i\mat{J}^{RFI^{H}}_{i,q}.
\end{equation}
These expressions can be arranged as matrix products,

\begin{equation}
\mat{V}^{RFI}_{pq} = \mat{J}^{RFI}_p \mat{C}^{RFI} \mat{J}^{RFI^{H}}_q,
\end{equation}
\begin{equation}
\mat{V}^{RFI} = \mat{J}^{RFI}\mat{C}^{RFI}\mat{J}^{RFI^{H}}, \quad \mat{J}^{RFI} = \begin{bmatrix} \mat{J}^{RFI}_1 \\ \vdots \\ \mat{J}^{RFI}_P\end{bmatrix},
\end{equation}
with 
$\mat{J}^{RFI}_{p} = \begin{bmatrix} \mat{J}^{RFI}_{1,p}& \hdots & \mat{J}^{RFI}_{D,p}\end{bmatrix}$
and $\mat{C}^{RFI} = \mbox{diag}(\mat{C}^{RFI}_1, \hdots,  \mat{C}^{RFI}_D)$

Let us note $\mat{R}^{RFI}$ the sample covariance estimate associated to $\mat{V}^{RFI}$. 
By definition $\mat{R}^{RFI}$ follows a Wishart distribution of degree $N$ and mean $\mat{V}^{RFI} = \mat{A}\mat{A}^H$,
$\mat{A} = \mat{J}^{RFI}\mat{C}^{RFI^{1/2}}$, and can be written

\begin{equation}
\mat{R}^{RFI} =  \mat{A}\mat{M}\mat{A}^H, \quad \mat{M}\sim \mathcal{W}(N,\mat{I}).
\end{equation}
For $(p,q) \in \mathcal{B}$, the visibility matrix for the baseline formed by the $p^{th}$ and $q^{th}$ sensors can be expressed as,

\begin{equation}
\mat{R}^{RFI}_{pq} = \mat{A}_p\mat{M}\mat{A}_q^H, \quad \mat{A}_p =  \mat{J}^{RFI}_{p}\mat{C}^{RFI^{1/2}}.
\end{equation}
Applying the vec operation,
\begin{equation}
\vec{r}^{RFI}_{pq} = \mbox{vec}(\mat{R}^{RFI}_{pq}) = (\mat{A}_q^* \otimes \mat{A}_p) \mbox{vec}(\mat{M}).
\end{equation}
Finally, all the visibilities induced by the RFI are concatenated in a vector to obtain the low-rank structure 

\begin{equation}\label{RFI-visibility}
\vec{r}^{RFI} = \mat{W}\mbox{vec}(\mat{M}),
\ \mbox{with} \ \mat{W} = \begin{bmatrix} \mat{A}_{2}^* \otimes \mat{A}_1 \\ \vdots \\ \mat{A}_{N}^* \otimes \mat{A}_{N-1} \end{bmatrix} 
\mbox{and} \ \mat{M}\sim \mathcal{W}(N, \mat{I}).
\end{equation}
As the number of samples increases, $\mat{M}$ appears to be closer to the asymptotic distribution of Wishart which is a Generalized Complex Normal distribution \cite{mahot2013} 
\begin{equation}
\sqrt{N}\mbox{vec}(\mat{M} - \mat{I}) \overset{d}{\rightarrow} \mathcal{GCN}(\mat{0}, \mat{I}, \mat{I}).
\end{equation}

Consequently, when considering sufficient samples so that the central limit theorem can be applied, equation
(\ref{eqn:model-no-outliers}) can be modified in (\ref{eqn:model-rfi-monofreq}) to accurately model the presence of RFI.
The maximum value for the rank of $\mat{W}$ can be identified from equation (\ref{RFI-visibility}), $\rho_{max}= 4L^2$. 

\subsection{Multi-frequency model}

In the previous sections, we only considered one frequency channel. Let us consider measurements across $F$ frequency channels.
Following the work presented in \cite{Yatawatta15}\cite{Brossard17}, we model the frequency dependance of the Jones matrices by 
a polynom. Considering the $f^{th}$ frequency channel, the $p^{th}$ receiver and the $i^{th}$ source, the Jones matrix $\mat{J}_{i,p}(f)$ is 
expressed as
\begin{equation}
  \mat{J}_{i,p}(f) = \sum_{k=1}^K b_k(f) \mat{Z}_{i,p,k}, \quad b_k(f) = \left(\frac{f-f_0}{f_0}\right)^{k-1},
\end{equation}
where $K$ is the polynomial order and $f_0$ the central frequency.
The expression is simplified using $\mat{B}_f = [b_1(f), \hdots, b_K(f)] \otimes \mat{I}_2$ ,

\begin{equation}
\mat{J}_{i,p}(f) = \mat{B}_f\mat{Z}_{i,p}, \quad \mat{Z}_{i,p} = \begin{bmatrix} \mat{Z}_{i,p,1} \\ \vdots \\ \mat{Z}_{i,p,K} \end{bmatrix}.
\end{equation}
The block matrix containing all the polynomial coefficient is noted  $\Z$. 
The ideal visibility vector for the $f^{th}$ frequency channel and the baseline formed by the $p^{th}$ and $q^{th}$ antenna
becomes
\begin{equation}
\vec{v}_{f,pq}(\Z) = \sum_{i=1}^D ((\mat{B}_f^* \mat{Z}_{i,q}^*) \otimes (\mat{B}_f \mat{Z}_{i,p}) )\mbox{vec}(\mat{C}_i)
\quad \mbox{and} \ 
\vec{v}_f(\Z) = \begin{bmatrix} \vec{v}_{f, 12}(\Z) \\ \vdots \\ \vec{v}_{f,(P-1)P}(\Z)\end{bmatrix}.
\end{equation}

The model expressed in (\ref{eqn:model-rfi-monofreq}) can be 
written using the formalism of mixed effect model to take into account the diversity accross frequency channels.
The presence of RFI in a given frequency channel as well as the thermal noise
associated to the frequency channel are considered as random effects and modeled
by $\mat{W}_f\vec{y}_f + \vec{n}_f$. On the other hand, the contribution of the calibrator sources to the visibilities, 
which frequency behavior is assumed known, is considered as fixed effect. The resulting model can be written as, 
\begin{equation}\label{eqn:model-freq}
\vec{r}_f= \vec{v}_f(\Z)  + \mat{W}_f\vec{y}_f + \vec{n}_f,
\quad
\vec{y}_f \sim \mathcal{CN}(\mbox{vec}(\mat{I}), \mat{I}) \ \mbox{and} \ \vec{n}_f \sim \mathcal{CN}(\vec{0}, \sigma^2\mat{I})
\end{equation}
% with $\rho \leq 4F_{RFI}L^2$, $F_{RFI}$ being the number of frequency band affected by RFI.

Ideally, since RFI can be found on multiple frequency channels, a low-rank matrix $\mat{W}_f$ would be assigned to each frequency channel affected by RFI.
To ease computation, we reduce the model dimension by considering a common low-rank matrix $\mat{W}$ that englobes the RFI structures for all the frequency channels.
Thus, the model is factorized in a simpler mixed effect model as $\mat{W}_f = \sigma_f \mat{W}$, where $\mat{W}$ is shared across all frequencies (absorbing the RFI structure) and
$\sigma_f$ is a soft weight. 

% The RFI matrix $\mat{W}$ is then weighted 
% by a soft activation $\sigma_f$ which value depends on the frequency channel.

By concatenating the measured \rev{visibilities} for all the frequency channels, we deduce the proposed model for RFI corrupted radio interferometric data.
\begin{equation}\label{eqn:model}
\vec{r}= \vec{v}(\Z)  + \left((\sigma_1, \hdots , \sigma_F) \otimes \mat{W} \right)\vec{y} + \vec{n},
\end{equation}
\rev{
with $\mat{W} \in \mathbb{C}^{4N_{\mathcal{B}} \times M}$,  $\vec{y}=[\vec{y}_1, \hdots , \vec{y}_F]$, $\forall{f} \in [1,F], \vec{y}_f \sim \mathcal{CN}(\mbox{vec}(\mat{I}), \mat{I})$ and
$\vec{n} \sim \mathcal{CN}(\vec{0}, \sigma^2\mat{I})$. 
The number of columns, M, of the tall matrix, $\W$, corresponds to its maximal rank (and also the maximal rank of the matrix $\W\W^H$),
so is set in order to ensure $\mbox{rank}(\W) \leq M \leq 4 N_B$.}
%The maximum value for the rank of $\mat{W}$ is $\rho_{max} = 4N_{\mathcal{B}}$. 
% However if there is some redundancy in the
% RFI structure for several frequency channels some column of $\mat{W}$ will be correlated, resulting in 
% a lower effective rank for $\mat{W}$. %Il faut approximativement P> 16L pour assurer rang faible.

\section{Maximum likelihood estimation using a SAGE algorithm}

In this section, we present an algorithm that aims at estimating all the unknown parameters of the model  
specified by (\ref{eqn:model}). The set of parameters is
\begin{equation}
\vec{\Theta} = \{ \mat{Z}, \mat{W}, \sigma^2, \sigma_1, \hdots , \sigma_F\}
\end{equation}
The parameter of interest for radio interferometer calibration is the set of polynomial coefficients $\Z$, used to
calibrate the measurements.
Given a data vector $\vec{r}\in\mathcal{C}^{1\times 4N_{\mathcal{B}}F}$ of measured visibilities, the log-likelihood for
the model specified by (\ref{eqn:model}) is 

\begin{equation}\label{eqn:log-lik}
  \begin{aligned}
    \mathcal{L}(\vec{r}, \vec{\theta}) &= -\log\left(\det{\sigma^2\mat{I} + \mat{\Psi}\mat{\Psi}^H}\right) \\ &- 
      (\vec{r} - \vec{v}(\Z)- \mat{\Psi}({\rm vec}(\eye{4N_{\mathcal{B}}}) \otimes \mathbf{1}_F))(\sigma^2\mat{I} + \mat{\Psi}\mat{\Psi}^H)^{-1}(\vec{r} - \vec{v}(\Z) - \mat{\Psi}({\rm vec}(\eye{4N_{\mathcal{B}}}) \otimes \mathbf{1}_F))^H
  \end{aligned}
\end{equation}
where $\mat{\Psi} = (\sigma_1, ..., \sigma_F) \otimes \mat{W}$ and $\vec{\theta} = \left[\sigma^2, \mbox{vec}(\mat{Z}), \mbox{vec}(\mat{W}), \sigma_1, ..., \sigma_F  \right]$, and $\mathbf{1}_F = (1, \dots, 1)$.

Direct maximization of the likelihood is not tractable, thus we propose to use a SAGE based algorithm to
find the Maximum Likelihood Estimate (MLE) for the proposed model
\cite{Fessler94} \cite{KS00}.
The SAGE is a variant of the Expectation-Maximization (EM) algorithm.
In its simplest form, the EM algorithm is an iterative approach to the computation of the MLE when observations can be viewed as
incomplete data.
There has been sufficient work on the convergence properties of the EM algorithm, though it is 
known that an EM procedure increases the likelihood at each iteration converging to at least 
a local maxima \cite{Dempster77}\cite{Wu83}.
Yet, the EM algorithm is known for its slow convergence and possible difficult maximization step
when the space of parameters is of high dimensions \cite{wang15}. The SAGE algorithm 
proposed in \cite{Fessler94} tackles those problems by using alternating hidden data spaces to sequentially
update the parameter vector. 
\rev{
In \cite{Kazemi11}, the SAGE algorithm is used to separate signals along different directions while 
assuming a gaussian noise on the visibilities, whereas, in this paper, the SAGE algorithm is used to separate
celestial and RFI signals with a model that specifically considers the presence of RFI.
}
In practice, the SAGE consists in defining multiple spaces of parameters and 
their corresponding complete data spaces in which EM procedures are performed. 

Thereby, we define two sets of complete data from which we derive a SAGE procedure. 
For each hidden data space we compute the expectation of the log-likelihood of the complete data given the observed
incomplete data in the \textit{E-step} as a surrogate function that is maximized in the \textit{M-step}. Thus 
the set of complete data is defined so that the likelihood of its associated parameters leads to closed form maximizations.
The purpose of \rev{the proposed algorithm} is to estimate the RFI subspace ($\mat{W}$ and $(\sigma_f)_{1\leq f\leq F}$) 
with the first hidden data space and to estimate the calibration parameters ($\mat{Z}$ and $\sigma^2$) with the second hidden data 
space. 

The first set of complete data
contains the measured visibilities as well as the RFI subspace,
\begin{equation}
\mathcal{X}_1 = \{\vec{r}, \vec{y}\}.
\end{equation}
Its associated parameter vector defines the RFI subspace, $\vec{\theta}_1 = \left[\mbox{vec}(\mat{W}) , \sigma_1, ..., \sigma_F\right]$.
We choose the second set of complete
data to be the source contributions to the measured visibilities and their respective thermal noise, 
\begin{equation}\label{complete-set-2}
\mathcal{X}_2 = \{ (\vec{u}_i)_{1\leq i\leq D} :  \vec{u} = \vec{r} - \mat{\Psi}\vec{y} = \sum_{i=1}^D \vec{u}_i\},
\end{equation}
where 
$\vec{u}_i = \vec{v}_i(\Z) + \vec{n}_i$ with $\vec{v}_i(\Z)$ being the $i^{th}$ source \rev{visibilities} and 
$\vec{n}_i \sim \mathcal{CN}(0,\beta_i \sigma^2\mat{I})$ st $\sum_{i=1}^D \beta_i=1$ and $\beta_i>0$. 
%This set of complete data is close to the
%ones classicaly used to perform calibration with an EM \cite{Kazemi11}\cite{Ollier17}.
%We propose to take advantage of the proven relevance of such hidden latent space while taking into account the presence of RFI.
The calibration parameters,
$\vec{\theta}_2 = \left[\mbox{vec}(\mat{Z}), \sigma^2\right]$, are updated using this second set of complete data.
The corresponding surrogate functions can be expressed as,
\begin{equation}
    \begin{aligned}
Q_1(\vec{\theta}_1, \vec{\theta}_1^{(m)}) &=
  \Esp{\mathcal{X}_1 \mid \vec{r} ; \vec{\theta}_1^{(m)}}{\log(p(\vec{r}, \vec{y}; \vec{\theta}_1))},
\\ 
Q_{2}(\vec{\theta}_2, \vec{\theta}_2^{(m)}) &= \sum_{i=1}^D
  \Esp{\mathcal{X}_2 \mid \vec{r} ; \vec{\theta}_1^{(m)}}{\log(p(\vec{u}_i; \vec{\theta}_2))}.
    \end{aligned}
\end{equation}

\begin{proposition}
  The surrogate function $Q_1$ reduces to 
  \rev{
\begin{equation} \label{enq:Q1}
    \begin{aligned}
      Q_1(\vec{\theta}\vert \vec{\theta}^{(m)})  &= A
       - \log\mid \sigma^2 \mat{I}_{4FN_{\mathcal{B}}}\mid  
      \\ &-
      \sum_{f=1}^F \bigg( \frac{1}{\sigma^2}\lVert \vec{r}_f-\vec{v}_f - \sigma_f\mat{W}\yhat_f\rVert_2^2
      +
      \frac{\sigma_f^2}{\sigma^2}\mbox{Tr}(\mat{W}^H\mat{W}\Sigmahat_f) + \mbox{Tr}(\Sigmahat_f) 
      \\
      &+ \norm{\yhat_f - \mbox{vec}(\eye{})}_2^2 \bigg)
    \end{aligned}  
\end{equation}
  where A is a constant independant of the model parameters and,}
\begin{equation}\label{eqn:E-step-1}
    \begin{aligned}
    \vec{\hat{y}}_f
    &=
    \mbox{vec}(\mat{I}) + 
    \sigma_f^{(m)}\mat{W}^{(m)^{H}}(\sigma^{(m)^2} \mat{I} + \sigma_f^{(m)^{2}}\mat{W}^{(m)}\mat{W}^{(m)^{H}})^{-1}
    (\vec{r}_f - \vec{v}_f(\Z^{(m)}) - \sigma_f \mat{W} \mbox{vec}(\eye{})),
      \\
      \Sigmahat_f
      &=
      \mat{I} -
      \sigma_f^{(m)}\mat{W}^{(m)^{H}}(\sigma^{(m)^{2}} \mat{I} + \sigma_f^{(m)^{2}}\mat{W}^{(m)}\mat{W}^{(m)^{H}})^{-1}\sigma_f^{(m)}\mat{W}^{(m)^{H}}.
    \end{aligned} 
\end{equation}
\end{proposition}
\begin{proof}
  : see Appendix A.
\end{proof}
Maximization of $Q_1$ is done using Wirtinger derivatives \cite{hjorougnes07} and leads to the following \rev{closed-form} expressions (see Appendix B)

\begin{equation}\label{eqn:M-step-1-sig_f}
\begin{aligned}
\sigma_f^{(m+1)}
&=
\frac{\Re\bigg((\vec{r}_f-\v_f)^H\mat{W}\vec{\hat{y}}_f\bigg)}{\mbox{Tr}\big(\mat{W}^H\mat{W}(\Sigmahat_f + \vec{\hat{y}}_f\vec{\hat{y}}_f^H) \big)},
\end{aligned}
\end{equation}

\begin{equation}\label{eqn:M-step-1-W}
\begin{aligned}
\mat{W}^{(m+1)}
&=
\bigg(\sum_{f=1}^F \sigma_f \left(\vec{\hat{y}}_f(\vec{r}_f-\vec{v}_f)^H \right)^H \bigg) \bigg(\sum_{f=1}^F \sigma_f^2 \left(\Sigmahat_f +\vec{\hat{y}}_f\vec{\hat{y}}_f^H\right)^H \bigg)^{-1}.
\end{aligned}
\end{equation}

\rev{
It is worth mentioning that there is an ambiguity between the updates of $\W$ and $\sigma_1, \dots, \sigma_F$.
In order to avoid any ambiguity in the estimation of $\W$ and $(\sigma_1, \hdots, \sigma_f)$, 
we choose to impose a unit norm to $\W$, $||\W||_2 = 1$,

%\begin{equation}
\begin{align}
  \sigma_f^{(m+1)} &= \norm{\mat{W}^{(m+1)}}_2 \times \sigma_f^{(m+1)}
   \label{normsigmaf}
  \\
  \mat{W}^{(m+1)}
  &=
  \frac{\mat{W}^{(m+1)}}{\norm{\mat{W}^{(m+1)}}_2}.
  \label{normW}
\end{align}
%\end{equation}
}

\begin{proposition}
  The surrogate function $Q_2$ reduces to
\begin{equation}\label{eqn:Q2}
  \begin{aligned}
    Q_2(\vec{\theta}\vert \vec{\theta}^{(m)}) 
    &=
      \sum_{i=1}^D C_{u_{i}} - \log\mid \beta_i \sigma^2\mat{I}_{4FN_{\mathcal{B}}}\mid  - \frac{1}{\beta_i \sigma^2} \norm{\uihat - \v_i(\Z)}_F^2 - \frac{1}{\beta_i \sigma^2} \mbox{Tr}(\hat{\mat{\Sigma}}_{{\vec{u}}_i}),
  \end{aligned}
  \end{equation}
\rev{where $C_{u_{i}}$ is a constant independant of the model parameters and,}
\begin{equation}\label{eqn:E-step-2}
    \begin{aligned}
    \hat{\vec{u}}_i
    =
    \vec{v}_i(\Z^{(m)}) + \beta_i \sigma^2 (\sigma^2\mat{I} + \mat{\Psi}^{(m)}\mat{\Psi}^{(m)^{H}})^{-1}(\vec{r} - \vec{v}(\Z^{(m)})- \mat{\Psi}(\mbox{vec}(\eye{}) \otimes \mathbf{1}_F ))
\\
    \hat{\mat{\Sigma}}_{{\vec{u}}_i}
      =
        \beta_i\sigma^2 \mat{I}_{4FN_{\mathcal{B}}} - \beta_i^2 \sigma^4 (\sigma^2\mat{I} + \mat{\Psi}^{(m)}\mat{\Psi}^{(m)^{H}})^{-1}.
    \end{aligned}
  \end{equation}
\end{proposition}

\begin{proof}
  : see Appendix C.
\end{proof}
Maximization of (\ref{eqn:Q2}) using Wirtinger derivatives leads to the following expression for $\sigma^2$ (see Appendix D),
  \begin{equation}\label{eqn:M-Step-2-sigma2}
      \sigma^{2^{(m+1)}}
      =
      \frac{1}{4FN_{\mathcal{B}}D}  \sum_{i=1}^D \frac{1}{\beta_i}\left( \norm{\hat{\vec{u}}_i  - \v_i(\Z)}_F^2 + \mbox{Tr}(\hat{\mat{\Sigma}}_{{\vec{u}}_i}) \right).
  \end{equation}
$Q_2$ is maximized over $\Z$ with a block coordinate descent \cite{FHHT07}, leading to the following update (see Appendix D),
\begin{equation}\label{eqn:M-Step-2-Z}
  \mbox{vec}(\Z^{(m+1)^*}_{i,p}) = (\mat{M}_{i,p} + \tilde{\mat{M}}_{i,p})^{-1}\mbox{vec}(\mat{T}_{i,p}), 
\end{equation}
with 
$
\mat{M}_{i,p} = \sum\limits_{f=1}^F \sum\limits_{q=p+1}^{N}
\mat{F}_{i,f,q}^T\mat{F}_{i,f,q}^* \otimes  \mat{B}_f^T \mat{B}_f^*,
$
$
\tilde{\mat{M}}_{i,p} = \sum\limits_{f=1}^F \sum\limits_{q=1}^{p-1}
\mat{G}_{i,f,q}^T\mat{G}_{i,f,q}^* \otimes  \mat{B}_f^T \mat{B}_f^*,
$
and \\
$
\mat{T}_{i,p} = \sum\limits_{f=1}^F \sum\limits_{q=p+1}^{N} \mat{B}_f^T \hat{\mat{U}}_{i,f,pq}^*\mat{F}_{i,f,q}
+
\sum\limits_{f=1}^F \sum\limits_{q=1}^{p-1} \mat{B}_f^T \hat{\mat{U}}_{i,f,qp}^T\mat{G}_{i,f,q},
$
where
$\mat{F}_{i,f,q} = \mat{B}_f^* \Z_{i,q}^{(m+1)^*} \mat{C}_{i,f}^T$
and \\
$\mat{G}_{i,f,q} = \mat{B}_f^* \Z_{i,q}^{(m+1)^*} \mat{C}_{i,f}^*.$

The SAGE algorithm consists in applying the two SAGE steps iteratively until convergence. We consider that convergence is
achieved when the variation of the log-likelihood between two iterations is less than a chosen threshold $\epsilon>0$.

\begin{algorithm}[H]
    \textbf{Input:} $\epsilon$, $N_{sources}$, $N_{sensors}$, $N_{frequencies}$, $\vec{v}$, $\vec{r}$,  $\sigma^{2^{(0)}}$, $\Z^{(0)}$, $\mat{W}^{(0)}$, $\sigma_1^{(0)}, ...,\sigma_F^{(0)}$.\newline
    \textbf{Output:} $\sigma^2$, $\Z$, $\mat{W}$, $\sigma_1$, ..., $\sigma_f$
    \caption{SAGE algorithm for multi-frequency calibration in presence of RFI}\label{sage}
      \begin{algorithmic}[1]
        \While{ ($\mathcal{L}(\vec{r},\vec{\theta}^{(m+1)}) -  \mathcal{L}(\vec{r},\vec{\theta}^{(m)})) \geq \epsilon$}
          \State Update $\hat{\vec{y}}, \hat{\mat{\Sigma}}_{\vec{y}}$ using (\ref{eqn:E-step-1})
          \For{$f$ in $[1:N_{frequencies}]$}
            \State Update $\sigma_f$ using (\ref{eqn:M-step-1-sig_f})
          \EndFor
          \State Update $\mat{W}$ using (\ref{eqn:M-step-1-W})
          \rev{
          \State Update $\sigma_1, \dots, \sigma_F$ using (\ref{normsigmaf})
          \State Update $\mat{W}$ using (\ref{normW})
          }
          \For{$i$ in $[1:N_{sources}]$}
            \State Update $\hat{\vec{u}}_i, \hat{\mat{\Sigma}}_{{\vec{u}}_i}$ using (\ref{eqn:E-step-2})
            \For{$p$ in $[1:N_{sensors}]$}
                \State Update $\Z[i, p]$ using (\ref{eqn:M-Step-2-Z})
            \EndFor
          \EndFor
          \State Update $\sigma^2$ using (\ref{eqn:M-Step-2-sigma2})
        \EndWhile
        \State {\textbf{Return:} ($\sigma^2$, $\Z$, $\mat{W}$, $\sigma_1$, ..., $\sigma_f$)}

      \end{algorithmic}
\end{algorithm}
\label{alg_4}

\section{Numerical results}
\rev{
In this section, several simulations are presented to investigate the 
robustness of the proposed model and to study the performance of the 
proposed algorithm.
To that end, we consider a radio interferometer of $P=8$ receivers 
observing a sky composed of $D=2$ calibrator sources along $F=32$ 
frequency channels. The frequency channels and the central frequency 
are chosen such that $\frac{f-f_0}{f_0}\in[-1,1]$.
The sky is composed of two unpolarized calibrator sources of flux 
100Jy and 50Jy. The matrix $\Z$ is generated using a uniform 
distribution, for $i\in[1,D]$, $p\in[1,P]$, $\Re(\Z_{i,p}) \sim 
\mathcal{U}(0,1)$ and $\Im(\Z_{i,p}) \sim \mathcal{U}(0,1)$.
The maximum number of iterations for the various iterative algorithms used is set to 
15 and 100 Monte Carlo repetitions are performed for a thermal noise SNR of 15dB.
We compute for each run the Normalized Mean Square Error (NMSE) for the parameter of interest, 
$\mbox{NMSE} = \frac{1}{100} \sum_{j=1}^{100} \frac{\norm{\Z - \hat{\Z}_j}_2^2}{\norm{\Z}_2^2}$.
We first show that our method outperforms standard projections methods \cite{LV00} in the context of strong RFI.
We then compare our algorithm to a method considering a robust noise model, namely the Student-t
distribution \cite{yatawatta2020stochastic}. Finally, we study the convergence 
rate and the influence of the rank of $\mat{W}$ for the proposed algorithm.

In the presence of strong RFI on specific frequency channels, it is possible to project out the 
RFI subspace by using the highest eigenvalues of the correlation matrix \cite{LV00}, reconstructing the
filtered visibilities free from RFI. Calibration can then be performed on the filtered visibilities.
We compare our proposed calibration algorithm that does not need to filter out the RFI to a classical calibration  
algorithm performed on filtered visibilities \cite{Kazemi11}.
A sky corrupted by 2 RFI, with Stokes parameter $ [100 \ 10 \ 50 \ 30]$ and $[50 \ 0\ 0 \ 0]$, 
is simulated.
\begin{figure}[H]
\centering
% This file was created with tikzplotlib v0.10.1.
\begin{tikzpicture}

\definecolor{darkgray176}{RGB}{176,176,176}
\definecolor{lightgray204}{RGB}{204,204,204}

\begin{axis}[
legend cell align={left},
legend style={
  fill opacity=0.8,
  draw opacity=1,
  text opacity=1,
  at={(0,0.75)},
  anchor=north west,
  draw=lightgray204
},
log basis y={10},
tick align=outside,
tick pos=left,
x grid style={darkgray176},
xtick={1,2,3,4},
xmin=0.8, xmax=4.2,
xtick style={color=black},
y grid style={darkgray176},
ymin=0.0184995611408439, ymax=1.18635038219195,
ymode=log,
ytick style={color=black},
ylabel=NMSE,
xlabel=Number of frequency channels affected
]
\addplot [semithick, black, mark=o, mark size=3, mark options={solid}]
table {%
1 0.738340849523572
2 0.766966289571952
3 0.826063057900216
4 0.981915843952361
};
\addlegendentry{projection + calibration algorithm} %\cite{Kazemi11}}
\addplot [semithick, blue, mark=o, mark size=3, mark options={solid}]
table {%
1 0.022351163355796
2 0.0347513410371096
3 0.0361755179748464
4 0.0443100310652846
};
\addlegendentry{proposed algorithm}
\end{axis}

\end{tikzpicture} 
\caption{Evolution of the NMSE of $\Z$ versus the number of neighboring frequency band affected by RFI}
\label{projection}
\end{figure}
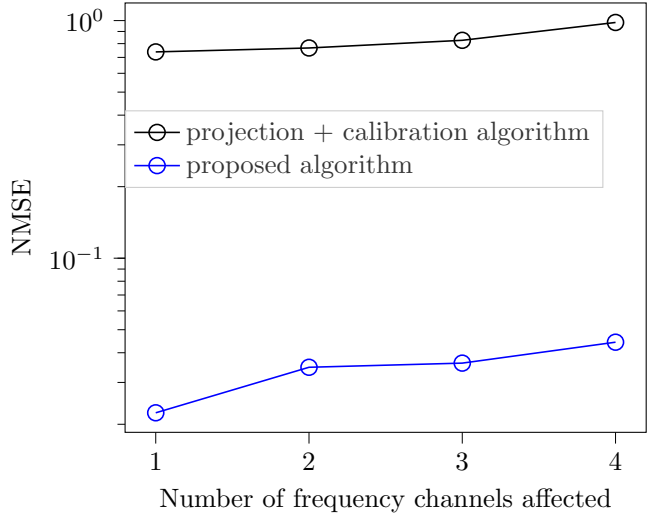
In figure \ref{projection} we plot the evolution of the NMSE for the parameter of interest versus the number of
frequency channels affected by
RFI. It appears that the proposed calibration algorithm leads to better performance. 
While filtering out the RFI, a part of the sources might be attenuated, leading to 
possible biased calibration solutions, whereas the proposed algorithm calibrates on the original
sources signals while considering the RFI subspace.

The Student-t distribution has proven itself to be a distribution of choice to
model the possible presence of outliers in the visibilities \cite{yatawatta2020stochastic}\cite{Ollier17}\cite{SBSKG20}.
Thus we compare
our proposed model to the model presented in \cite{yatawatta2020stochastic},
considering a Student-t distribution as additive noise on the visibilities.
Figure \ref{student1} shows the evolution of the NMSE versus the power of the RFI when only 
a few frequency channels are corrupted by interference.
RFI with a power that ranges from -10dB to 10dB are added on 
$10\%$ and $30\%$ of the frequency channels.
Figures \ref{student1}.b and \ref{student1}.d present the visibility amplitude across all the frequency bands for
one baseline when respectively $10\%$ and $30\%$ of the frequency channels are affected by RFI.
The proposed method displays, in figure \ref{student1}.a and \ref{student1}.c, performances that are similar to the state of the 
art methods, considering a Student-t distribution as additive noise, despite that such
methods are prone to good behavior in this specific context.
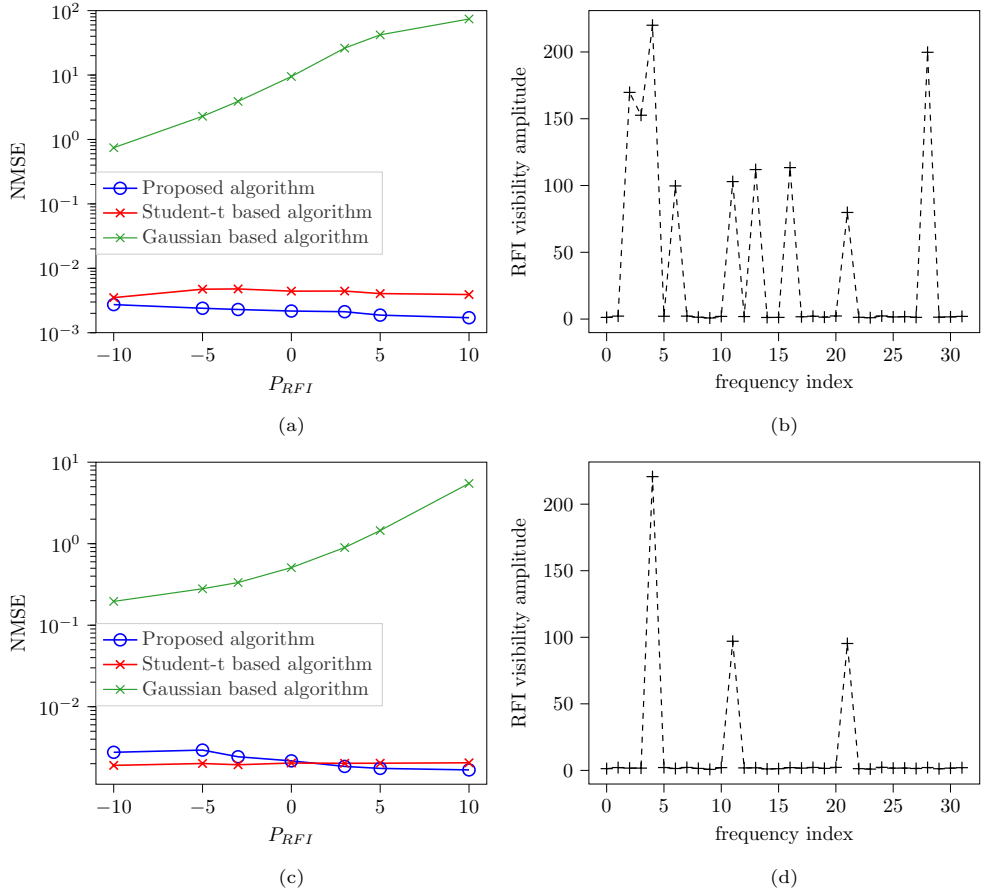
\begin{figure}[H]
  \begin{center}
  % This file was created with tikzplotlib v0.9.15.
\begin{tikzpicture}[scale=0.75]

\definecolor{color0}{rgb}{0.12156862745098,0.466666666666667,0.705882352941177}
\definecolor{color1}{rgb}{1,0.498039215686275,0.0549019607843137}
\definecolor{color2}{rgb}{0.172549019607843,0.627450980392157,0.172549019607843}

\begin{groupplot}[group style={group name=student1a, group size=2 by 1,}]
\nextgroupplot[
legend cell align={left},
legend style={
  fill opacity=0.8,
  draw opacity=1,
  text opacity=1,
  at={(0,0.5)},
  anchor=north west,
  draw=white!80!black
},
log basis y={10},
tick align=outside,
tick pos=left,
x grid style={white!69.0196078431373!black},
xlabel={$P_{RFI}$},
xmin=-11, xmax=11,
xtick style={color=black},
y grid style={white!69.0196078431373!black},
ylabel={NMSE},
ymin=0.00100263421315401, ymax=100,
ymode=log,
ytick style={color=black}
]
\addplot [thick, blue, mark=o, mark size=3, mark options={solid}]
table {%
-10 0.00273186831146587
-5 0.00239822057671407
-3 0.00229396072780927
0 0.00217129802114728
3 0.00212017038390782
5 0.00187572812897298
10 0.00171001625151454
};
\addlegendentry{Proposed algorithm}
\addplot [thick, red, mark=x, mark size=3, mark options={solid}]
table {%
-10 0.0035069895587282
-5 0.00473883301842333
-3 0.00478261642814548
0 0.00442010647551613
3 0.00443430240357264
5 0.00405184495555235
10 0.00391566360198724
};
\addlegendentry{Student-t based algorithm}
\addplot [semithick, color2, mark=x, mark size=3, mark options={solid}]
table {%
-10 0.744699885268762
-5 2.29047223721435
-3 3.89943714127782
0 9.48876555962248
3 26.1973390066951
5 42.0941221412566
10 74.1574167340658
};
\addlegendentry{Gaussian based algorithm}
\end{groupplot}
\node[below = 1cm of student1a c1r1.south] {\scriptsize (a)};
\end{tikzpicture}
  % This file was created with tikzplotlib v0.9.12.
\begin{tikzpicture}[scale=0.75]

\begin{axis}[
    legend cell align={left},
    legend style={fill opacity=0.8, draw opacity=1, text opacity=1, draw=white!80!black},
    tick align=outside,
    tick pos=left,
    x grid style={white!69.0196078431373!black},
    xlabel={frequency index},
    xmin=-1.55, xmax=32.55,
    xtick style={color=black},
    y grid style={white!69.0196078431373!black},
    ylabel={RFI visibility amplitude},
    ymin=-10.1450931061398, ymax=230.981322051647,
    ytick style={color=black},
    alias=ax
    ]
    \addplot [semithick, black, dashed, mark=+, mark size=3, mark options={solid}]
    table {%
    0 1.23604733859119
    1 2.2148096840292
    2 169.731462762191
    3 152.615562476599
    4 220.021030453566
    5 2.10558210343347
    6 99.6962138742889
    7 2.21107972581391
    8 1.40214378011963
    9 0.815198491941451
    10 1.98129412255346
    11 102.939060974549
    12 1.89076451350539
    13 111.948652659882
    14 1.19781288208262
    15 1.26089817778745
    16 113.413934467585
    17 1.72757692029306
    18 2.11063806837427
    19 1.43556200960145
    20 2.31642555813541
    21 79.8167115412725
    22 1.31710395615044
    23 1.01661212447712
    24 2.32170928787731
    25 1.58532218824354
    26 1.76528016096065
    27 1.28933353883308
    28 199.748073902575
    29 1.41058291495562
    30 1.63215013762547
    31 2.00515754941231
    };
    \end{axis}
    \node[below = 1cm of ax.south] {\scriptsize (b)};
    \end{tikzpicture}
        
  % This file was created with tikzplotlib v0.9.15.
\begin{tikzpicture}[scale=0.75]

\definecolor{color0}{rgb}{0.12156862745098,0.466666666666667,0.705882352941177}
\definecolor{color1}{rgb}{1,0.498039215686275,0.0549019607843137}
\definecolor{color2}{rgb}{0.172549019607843,0.627450980392157,0.172549019607843}

\begin{groupplot}[group style={group name=student1c, group size=2 by 1,}]
\nextgroupplot[
legend cell align={left},
legend style={
  fill opacity=0.8,
  draw opacity=1,
  text opacity=1,
  at={(0,0.5)},
  anchor=north west,
  draw=white!80!black
},
log basis y={10},
tick align=outside,
tick pos=left,
x grid style={white!69.0196078431373!black},
xlabel={$P_{RFI}$},
xmin=-11, xmax=11,
xtick style={color=black},
y grid style={white!69.0196078431373!black},
ylabel={NMSE},
ymin=0.00111969198175799, ymax=10,
ymode=log,
ytick style={color=black}
]
\addplot [thick, blue, mark=o, mark size=3, mark options={solid}]
table {%
-10 0.00275910567525563
-5 0.0029371880973932
-3 0.00243291237513639
0 0.00216590663939426
3 0.00185174039715675
5 0.00175106517855238
10 0.00167813146544122
};
\addlegendentry{Proposed algorithm}
\addplot [thick, red, mark=x, mark size=3, mark options={solid}]
table {%
-10 0.00190709398572769
-5 0.00200998554106704
-3 0.00194078703899377
0 0.0020382472552036
3 0.00202177414858793
5 0.00202652823836548
10 0.00205141692657592
};
\addlegendentry{Student-t based algorithm}
\addplot [semithick, color2, mark=x, mark size=3, mark options={solid}]
table {%
-10 0.195999760170588
-5 0.280163581546203
-3 0.334000175281299
0 0.508706491370145
3 0.897368617391991
5 1.45089856012847
10 5.48749615809708
};
\addlegendentry{Gaussian based algorithm}
\end{groupplot}
\node[below = 1cm of student1c c1r1.south] {\scriptsize (c)};
\end{tikzpicture}
  % This file was created with tikzplotlib v0.9.12.
\begin{tikzpicture}[scale=0.75]

\begin{axis}[
    legend cell align={left},
    legend style={fill opacity=0.8, draw opacity=1, text opacity=1, draw=white!80!black},
    tick align=outside,
    tick pos=left,
    x grid style={white!69.0196078431373!black},
    xlabel={frequency index},
    xmin=-1.55, xmax=32.55,
    xtick style={color=black},
    y grid style={white!69.0196078431373!black},
    ylabel={RFI visibility amplitude},
    ymin=-10.2439420245849, ymax=231.686228402213,
    ytick style={color=black},
    alias=ax
    ]
    \addplot [semithick, black, dashed, mark=+, mark size=3, mark options={solid}]
    table {%
    0 1.30374471116316
    1 2.12110298162905
    2 1.71216930751264
    3 1.70776419465182
    4 220.689402473722
    5 2.13712499124667
    6 1.41487206853869
    7 2.19681308165808
    8 1.46390143694307
    9 0.75288390390588
    10 2.06502539843802
    11 97.016012543503
    12 1.86450841598177
    13 1.92393138183724
    14 1.09259384779282
    15 1.26249148390807
    16 2.09206237155483
    17 1.70614688893173
    18 2.1983683139262
    19 1.51289891292974
    20 2.23408733322467
    21 95.3254470463247
    22 1.29366821209218
    23 0.980084571053343
    24 2.28153553992792
    25 1.6866288152975
    26 1.73293533557362
    27 1.41491072263256
    28 2.14391532388671
    29 1.24581957004545
    30 1.68921782537522
    31 2.03598819129523
    };
    \end{axis}
    \node[below = 1cm of ax.south] {\scriptsize (d)};
    \end{tikzpicture}
    
  \end{center}
  \caption{Evolution of the NMSE of $\Z$ versus the power of the strong RFI}
  \label{student1}
  \end{figure}
In reality, low RFI are predominant across multiple frequency
channels \cite{Fridman01}.
Subsequently, we simulated a sky perturbated by
low RFI (-15dB) on each frequency channel. Additionally, strong RFI are added
to $10\%$ of the frequency bands with a power that ranges from -10dB to 10dB as illustrated in
figure \ref{student2}.b for one baseline.
% Figure puissance RFI
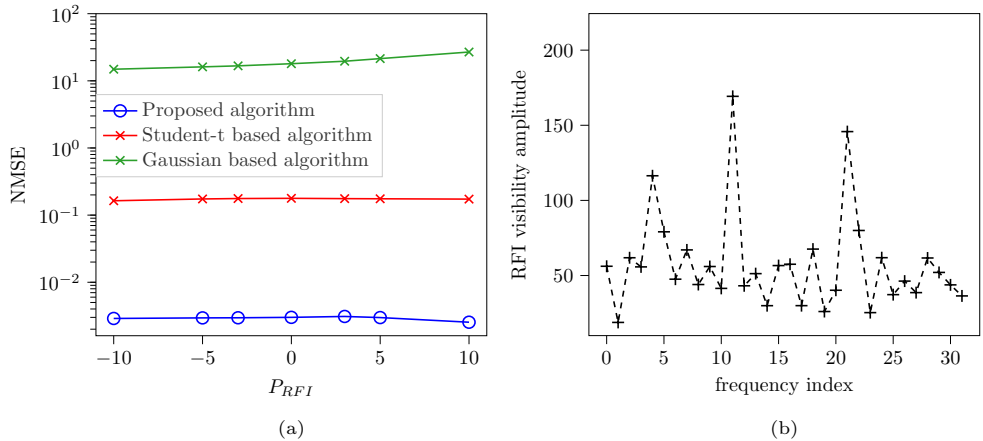
\begin{figure}[H] 
\begin{center}
  % This file was created with tikzplotlib v0.9.15.
\begin{tikzpicture}[scale=0.75]

\definecolor{color0}{rgb}{0.12156862745098,0.466666666666667,0.705882352941177}
\definecolor{color1}{rgb}{1,0.498039215686275,0.0549019607843137}
\definecolor{color2}{rgb}{0.172549019607843,0.627450980392157,0.172549019607843}

\begin{groupplot}[group style={group name=student2a, group size=1 by 1,}]
\nextgroupplot[
legend cell align={left},
legend style={
  fill opacity=0.8,
  draw opacity=1,
  text opacity=1,
  at={(0,0.75)},
  anchor=north west,
  draw=white!80!black
},
log basis y={10},
tick align=outside,
tick pos=left,
x grid style={white!69.0196078431373!black},
xlabel={$P_{RFI}$},
xmin=-11, xmax=11,
xtick style={color=black},
y grid style={white!69.0196078431373!black},
ylabel={NMSE},
ymin=0.00159855859826401, ymax=100,
ymode=log,
ytick style={color=black}
]
\addplot [thick, blue, mark=o, mark size=3, mark options={solid}]
table {%
-10 0.00288996457752521
-5 0.00294758962343013
-3 0.00295848320051037
0 0.00300192340632481
3 0.00309541100695489
5 0.00297922993461361
10 0.00254059770204755
};
\addlegendentry{Proposed algorithm}
\addplot [thick, red, mark=x, mark size=3, mark options={solid}]
table {%
-10 0.163537444976242
-5 0.174049172146694
-3 0.176578760802136
0 0.177843387096684
3 0.176000971172367
5 0.175120519766683
10 0.173135135335972
};
\addlegendentry{Student-t based algorithm}
\addplot [thick, color2, mark=x, mark size=3, mark options={solid}]
table {%
-10 14.8935615422962
-5 16.1423427337471
-3 16.7147409097984
0 17.9985548780135
3 19.5839468495351
5 21.4322940629914
10 26.8585586981163
};
\addlegendentry{Gaussian based algorithm}
\end{groupplot}
\node[below = 1cm of student2a c1r1.south] {\scriptsize (a)};
\end{tikzpicture}
  % This file was created with tikzplotlib v0.9.12.
\begin{tikzpicture}[scale=0.75]

\begin{axis}[
    legend cell align={left},
    legend style={fill opacity=0.8, draw opacity=1, text opacity=1, draw=white!80!black},
    tick align=outside,
    tick pos=left,
    x grid style={white!69.0196078431373!black},
    xlabel={frequency index},
    xmin=-1.55, xmax=32.55,
    xtick style={color=black},
    y grid style={white!69.0196078431373!black},
    ylabel={RFI visibility amplitude},
    ymin=10, ymax=224.324864934687,
    ytick style={color=black},
    alias=ax
    ]
    \addplot [thick, black, dashed, mark=+, mark size=3, mark options={solid}]
    table {%
    0 56.1574855749062
    1 18.7829479781699
    2 61.8299877595042
    3 55.7452749936933
    4 116.392745614743
    5 79.0854836910719
    6 47.5612462918243
    7 67.0442079036465
    8 44.0080894666745
    9 55.9948699764906
    10 41.4367990319346
    11 169.256596566042
    12 43.1617010733183
    13 51.2530161209205
    14 29.9259207940526
    15 56.6225672513205
    16 57.5043631040307
    17 29.9449836307278
    18 67.5890868400211
    19 26.0305219080151
    20 40.1942734976303
    21 145.789379688015
    22 79.9787426200356
    23 25.3048312366681
    24 61.8661759575574
    25 37.2716655286685
    26 46.2846016672575
    27 38.6167827867922
    28 61.6702899482086
    29 52.05693878197
    30 43.7787025373121
    31 36.4223986258917
    };
    \end{axis}
    \node[below = 1cm of ax.south] {\scriptsize (b)};
    \end{tikzpicture}
     
\end{center}
\caption{Evolution of the NMSE of $\Z$ versus the power of the strong RFI with additional low RFI on each frequency band}
\label{student2}
\end{figure}
We plot in figure \ref{student2}.a the evolution of the NMSE versus the power of the strong RFI.
A significant gain in performance can be noticed for the proposed algorithm compared to
the state-of-the-art. The contribution of our proposed model lies in taking into account the structure of the RFI,
adding robustness to the presence of RFI of various forms.
In figure \ref{rank}, the evolution of the NMSE for $\Z$ versus the chosen rank is displayed.
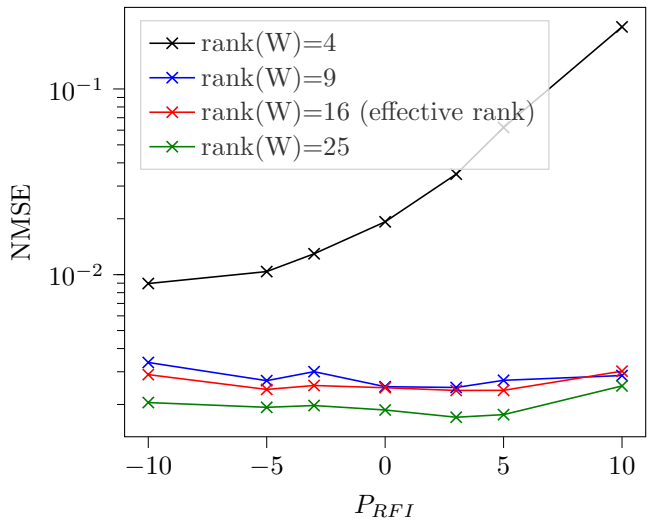
\begin{figure}[H]
\begin{center}
% This file was created with tikzplotlib v0.9.12.
\begin{tikzpicture}

\begin{axis}[
legend cell align={left},
legend style={
  fill opacity=0.8,
  draw opacity=1,
  text opacity=1,
  at={(0.03,0.97)},
  anchor=north west,
  draw=white!80!black
},
log basis y={10},
tick align=outside,
tick pos=left,
x grid style={white!69.0196078431373!black},
xlabel={$P_{RFI}$},
xmin=-11, xmax=11,
xtick style={color=black},
y grid style={white!69.0196078431373!black},
ylabel={NMSE},
ymin=0.00134015604042548, ymax=0.274941126615784,
ymode=log,
ytick style={color=black}
]
\addplot [semithick, black, mark=x, mark size=3, mark options={solid}]
table {%
-10 0.00895849519340547
-5 0.0103894363198136
-3 0.0129554400942707
0 0.0192581304815385
3 0.0347751441681139
5 0.0620530653275436
10 0.215846477309794
};
\addlegendentry{rank(W)=4}
\addplot [semithick, blue, mark=x, mark size=3, mark options={solid}]
table {%
-10 0.00336679262566035
-5 0.00268645866377849
-3 0.00300387537111792
0 0.00249319196203002
3 0.00246874515677111
5 0.00269834779632684
10 0.00286269672861398
};
\addlegendentry{rank(W)=9}
\addplot [semithick, red, mark=x, mark size=3, mark options={solid}]
table {%
-10 0.00289558685123236
-5 0.00240958942363874
-3 0.00253011957585437
0 0.00246117516057127
3 0.00237974643633799
5 0.00238316838916073
10 0.00302287568274536
};
\addlegendentry{rank(W)=16 (effective rank)}
\addplot [semithick, green!50.1960784313725!black, mark=x, mark size=3, mark options={solid}]
table {%
-10 0.00204813565890481
-5 0.00193070741161573
-3 0.00197375719952644
0 0.00186782715949625
3 0.0017070652075859
5 0.00176314085847633
10 0.00251662124462153
};
\addlegendentry{rank(W)=25}
\end{axis}

\end{tikzpicture}
\end{center} 
\caption{Evolution of the NMSE of $\Z$ versus the power of the RFI for multiple choice of the rank}
\label{rank}
\end{figure}
The effective rank used to simulate the visibilities is $\rho_{\rm{eff}}=16$. It can be noticed 
that a slight misjudgment of the rank does not change the performance of the algorithm.
These results illustrate the fact that the rank of $\W$ constrains the rank of the clutter noise covariance matrix, $\Phi = \W\W^H$.
If the rank of $\W$ is highly underestimated, the clutter noise covariance matrix will also have a rank highly underestimated, leading to poor reconstruction. 
Conversely, since there is no constraint in the rank of the estimated RFI structure matrix, $\W$, the clutter noise covariance
matrix can still be correctly estimated when the rank of $\W$ is overestimated.
This particular case might lead to overfitting in the estimate of $\W$. However, in practice,
we observe that our parameter of interest, $\Z$, is correctly estimated.
\begin{figure}[H]
  \begin{center}
    \input{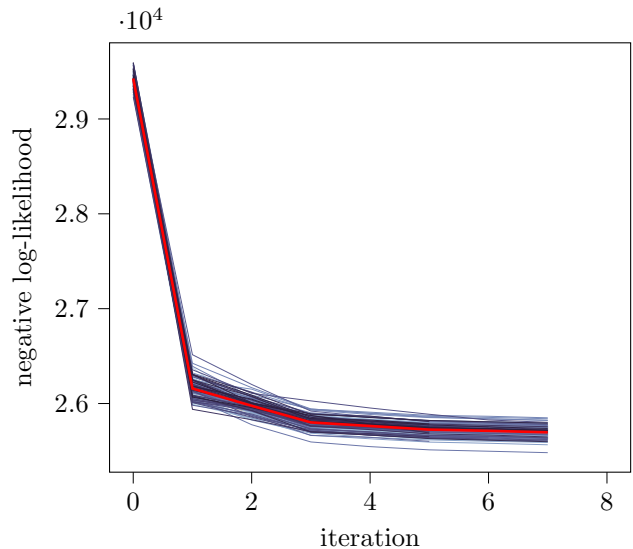}
  \end{center}
  \caption{Evolution of the negative log likelihood versus the numnber of iteration for 100 realizations}
  \label{convergence}
  \end{figure}
Figure \ref{convergence} presents the evolution of the negative log likelihood versus the 
number of iterations for 100 realizations. It appears that only a few iterations are needed for the proposed 
algorithm to converge, leading to a reasonable computational cost. 
}

\section{Conclusion}
\rev{
In this paper, we have first proposed a model for radio interferometer calibration that takes into account the presence of radiofrequency interferences, 
using a low-rank structured noise. The calibration parameters are obtained by a 
maximum likelihood estimation formulated thanks to a Space Alternating Generalized Expectation-Maximization algorithm. This latter leads to closed form updates for the model
parameters thanks to the choice of two judicious complete data sets. Numerical simulations show that the proposed estimator 
is robust to the presence of radiofrequency interferences and demonstrate a gain in performance compared to 
the state-of-the-art with reasonable computational time. Moreover, simulations 
show that a misjudgment of the rank of the design matrix $\mat{W}$ does not impact the performance of the algorithm.
}

%-----------------------------------------------------------------------------------------------------------------------------------------------%
%                                                                Appendix                                                                       %
%-----------------------------------------------------------------------------------------------------------------------------------------------%
\newpage
\appendix

\section{ Proof of Proposition 1}
We seek to compute 
$
Q_1(\vec{\theta}, \vec{\theta}^{(m)}) =
\Esp{\mathcal{X}_{c,1} \mid \vec{r} ; \vec{\theta}^{(m)}}{\log(p(\vec{r},\y ; \vec{\theta}))}.
$
Since $\vec{r}$ and $\vec{y}$ can be decomposed into mutually independant frequency vectors, $\vec{r} = [\vec{r}_1^T, ..., \vec{r}_F^T]$ and 
$\vec{y} = [\vec{y}_1^T, ..., \vec{y}_F^T]$ where
$\vec{r}_f = \vec{v}_f + \sigma_f\mat{W}\y_f + \vec{n}_f$, $Q_1$ can be 
expressed as 
\begin{equation}
  Q_1(\vec{\theta}, \vec{\theta}^{(m)}) = \sum_{f=1}^F
\Esp{\mathcal{X}_{c,1} \mid \vec{r} ; \vec{\theta}^{(m)}}{\log(p(\vec{r}_f,\y_f ; \vec{\theta}))}
\end{equation}

It is straightforward from the model definition that $\vec{r}_f \mid \y_f \sim \mathcal{CN}(\vec{v}_f(\Z) + \sigma_f\mat{W}\y, \ \sigma^2 \mat{I}_{4B})$ and $\y_f \sim \mathcal{CN}(\mbox{vec}(\mat{I}),\ \mat{I})$.
Thus the log-likelihoods of $\vec{r}_f \mid \vec{y}_f$ and $\vec{y}_f$ can respectively be written as
\begin{equation}
  \begin{aligned}
   \log(p(\vec{r}_f \mid \y_f ; \vec{\theta} )) &= \rev{-4FN_{\mathcal{B}}\log(\pi)} - \log\mid \sigma^2 \mat{I}\mid - \frac{1}{\sigma^2}(\vec{r}_f-\vec{v}_f(\Z) - \sigma_f\mat{W}\y)^H(\vec{r}_f-\vec{v}_f(\Z) - \sigma_f\mat{W}\y_f),\\
   \log(p(\y_f ; \vec{\theta})) &= \rev{-\rho\log(\pi)} - \log\mid \mat{I_\rho} \mid \ - \  \sum_{f=1}^F (\y_f - \mbox{vec}(\mat{I}) )^H (\y_f - \mbox{vec}(\mat{I})).
  \end{aligned}
\end{equation}
\def\Q{{Q_1(\vec{\theta} \mid \vec{\theta}^{(m)})}}
Thus, $\Q$ can be expressed as,
\begin{equation}
\begin{aligned}
% \Q &=
%   \Esp{\y \mid \x;\vec{\theta}^{(m)}}{\log \big(p\left(\x,\y ; \vec{\theta}\right) \big)}
% \\
\Q &= \sum_{f=1}^F
  \Esp{\y \mid \vec{r};\vec{\theta}^{(m)}}{\log(p(\vec{r}_f \mid \y_f ; \vec{\theta})) + \log(p(\y_f;\vec{\theta}))}
\\
 &=
\rev{A} + \sum_{f=1}^F - \log\mid \sigma^2 \mat{I}_{4B}\mid 
\\ &-
  \frac{1}{\sigma^2}\Esp{\y \mid \vec{r} ;\vec{\theta}^{(m)}}{(\vec{r}_f-\vec{v}_f(\Z) - \sigma_f\mat{W}\y_f)^H(\vec{r}_f-\vec{v}_f(\Z) - \sigma_f\mat{W}\y_f)}   
\\ & -\Esp{\y \mid \vec{r} ;\vec{\theta}^{(m)}}{(\y_f - \mbox{vec}(\mat{I}) )^H (\y_f - \mbox{vec}(\mat{I}))}
\\
 &=
\rev{A} - F\log\mid \sigma^2 \mat{I}_{4B}\mid 
+ \sum_{f=1}^F \frac{2\sigma_f}{\sigma^2}\Re((\vec{r}_f-\vec{v}_f)^H\mat{W}\Esp{\y \mid \vec{r};\vec{\theta}^{(m)}}{\y_f})
\\ &-
\frac{1}{\sigma^2}(\vec{r}_f-\vec{v}_f)^H(\vec{r}_f-\vec{v}_f)
- \frac{\sigma_f^2}{\sigma^2} \Esp{\y \mid \vec{r};\vec{\theta}^{(m)}}{\y_f^H\mat{W}^H\mat{W}\y_f}
\\ & -\Esp{\y \mid \vec{r} ;\vec{\theta}^{(m)}}{\y_f^H\y_f} + 2\mbox{vec}(\mat{I})^H\Esp{\y \mid \vec{r} ;\vec{\theta}^{(m)}}{\y_f}
 - \mbox{vec}(\mat{I})^H\mbox{vec}(\mat{I})
\\
% \Q &=
% C_1 + C_2 - \log\mid \sigma^2 \mat{I}_{4FN_{\mathcal{B}}}\mid -\ \log\mid \mat{I}_{\rho} \mid - \frac{1}{\sigma^2}(\x-\vec{v})^H(\x-\vec{v})
% \\ &+
% \frac{2}{\sigma^2}\Re((\x-\vec{v})^H\mat{\Psi}\yhat) - \frac{1}{\sigma^2}(\mbox{Tr}((\mat{\Psi}^H\mat{\Psi})\Sigmahat) + \yhat^H(\mat{\Psi}^H\mat{\Psi})\yhat)
% \\ &-
% \mbox{Tr}(\Sigmahat) - \yhat^H\yhat
% \\
% \Q &=
% C_1 + C_2 - \log\mid \sigma^2 \mat{I}_{4FN_{\mathcal{B}}}\mid -\ \log\mid \mat{I}_{\rho} \mid - \frac{1}{\sigma^2}(\x-\vec{v} - \mat{\Psi}\yhat)^H(\x-\vec{v} - \mat{\Psi}\yhat)
% \\ &-
% \frac{1}{\sigma^2}\mbox{Tr}((\mat{\Psi}^H\mat{\Psi})\Sigmahat) - \mbox{Tr}(\Sigmahat) - \yhat^H\yhat
\end{aligned}
\end{equation}
\rev{where A is a constant independant of the model parameters}
First the expression of the mean and covariance of $\vec{y}_f$ are derived, respectively 
$\yhat_f = \Esp{\y \mid \vec{r};\vec{\theta}^{(m)}}{\y_f}$ and 
$\Sigmahat_f = \Esp{\y \mid \vec{r};\vec{\theta}^{(m)}}{(\y_f-\yhat_f)(\y_f- \yhat_f)^H}$.
We consider $\vec{z} = [\y_f, \vec{r}_f]^T$, $\vec{z}$ being a multivariate Gaussian in which the 
mean of  $\y_f$ and $\vec{r}_f$ are respectively
\begin{equation}
  \vec{\mu}_{\y_f} = \mbox{vec}(\mat{I}), \quad \vec{\mu}_{\vec{r}_f} = \vec{v}_f(\Z) + \sigma_f \mat{W}\mbox{vec}(\eye{}).
\end{equation}
The Covariance \rev{block} matrices $\mat{\Sigma}_{\y\y}$, $\mat{\Sigma}_{\vec{r}\vec{r}}$ and $\mat{\Sigma}_{\vec{r}\y}$ are :

\begin{equation}
  \begin{aligned}
    \mat{\Sigma}_{\y\y} &= \Esp{}{(\y_f- \vec{\mu}_{\y_f})(\y_f - \vec{\mu}_{\y_f})^H} = \mat{I}\\
    \mat{\Sigma}_{\vec{r}\vec{r}} &= \Esp{\vec{r}}{(\vec{r}_f - \vec{\mu}_{\vec{r}_f})(\vec{r}_f - \vec{\mu}_{\vec{r}_f})^H} 
    = \sigma^{(m)^{2}} \mat{I}_{4B} + \sigma_f^{2^{(m)}}\mat{W}^{(m)}\mat{W}^{(m)^{H}} \\
    \mat{\Sigma}_{\y\vec{r}} &= \mat{\Sigma}_{\vec{r}\y}^H =
     \Esp{}{(\y_f - \vec{\mu}_{\y_f})(\vec{r}_f - \vec{\mu}_{\vec{r}_f})^H}\\
    &=\Esp{}{(\y_f - \mbox{vec}(\mat{I}) )(\sigma_f^{(m)}\mat{W}^{(m)}\vec{y}_f+ \vec{n}_f)^H}\\
    &=\Esp{}{\y_f(\sigma_f^{(m)}\mat{W}^{(m)}\vec{y}_f)^H} + \Esp{}{\y_f\vec{n}_f^H} - \mbox{vec}(\mat{I})(\sigma_f^{(m)}\mat{W}^{(m)}\Esp{}{\y_f})^H - \mbox{vec}(\mat{I})\Esp{}{\vec{n}_f^H}\\
    \end{aligned}
\end{equation}
Since $\y_f$ and $\vec{n}_f$ are considered independently distributed and $\Esp{}{\vec{n}_f} = \vec{0}$, $\Esp{}{\y_f\vec{n}_f^H} = \vec{0}$ and 
\begin{equation}
    \begin{aligned}
     \mat{\Sigma}_{\y\vec{r}} = \mat{\Sigma}_{\vec{r}\y}^H 
     &=
      \Esp{}{\y_f(\sigma_f^{(m)}\mat{W}^{(m)}\vec{y}_f)^H} - \mbox{vec}(\mat{I})(\sigma_f^{(m)}\mat{W}^{(m)}\Esp{}{\y_f})^H\\
    &= (\Esp{}{\y_f\y_f^H} - \mbox{vec}(\mat{I})\mbox{vec}(\mat{I})^H )\sigma_f^{(m)}\mat{W}^{(m)^H} = \sigma_f^{(m)}\mat{W}^{(m)^{H}}\\
    \end{aligned}
\end{equation}
Using \cite[p36]{Anderson03} the analytical expression for $\Sigmahat$ and $\yhat$ are deduced,
\begin{equation}
  \begin{aligned}
    \yhat_f &= \vec{\mu}_{\y_f} + \mat{\Sigma}_{\y\vec{r}}\mat{\Sigma}_{\vec{r}\vec{r}}^{-1}(\vec{r}_f - \vec{\mu}_{\vec{r}_f}) 
    \\&=
    \mbox{vec}(\mat{I}) +  \sigma_f^{(m)}\mat{W}^{(m)^{H}}(\sigma^{(m)^2} \mat{I}_{4B} + \sigma_f^{2^{(m)}}\mat{W}^{(m)}\mat{W}^{(m)^{H}})^{-1}
    (\vec{r}_f - \vec{v}_f(\Z^{(m)}) -  \sigma_f \mat{W}\mbox{vec}(\eye{}))
    \\
    \Sigmahat_f &= \mat{\Sigma}_{\y\y} - \mat{\Sigma}_{\y\vec{r}}\mat{\Sigma}_{\vec{r}\vec{r}}^{-1}\mat{\Sigma}_{\vec{r}\y} 
    \\& = \mat{I} -
    \sigma_f^{(m)}\mat{W}^{(m)^H}(\sigma^{(m)^{2}} \mat{I}_{4B} + \sigma_f^{2^{(m)}}\mat{W}^{(m)}\mat{W}^{(m)^{H}})^{-1}\sigma_f\mat{W}^{(m)}.
  \end{aligned}
\end{equation}

$\Esp{\y \mid \vec{r} ;\vec{\theta}^{(m)}}{\y_f^H\y_f}$ and $\Esp{\y \mid \vec{r} ;\vec{\theta}^{(m)}}{\y_f^H(\sigma_f^2\mat{W}^H\mat{W})\y_f}$ are then derived,
\begin{equation}
  \begin{aligned}
      \Esp{\y \mid \vec{r} ;\vec{\theta}^{(m)}}{\y_f^H\y_f} & = \Esp{\y \mid \vec{r} ;\vec{\theta}^{(m)}}{\mbox{Tr}(\y_f\y_f^H)}
      = \mbox{Tr}(\Esp{\y \mid \vec{r} ;\vec{\theta}^{(m)}}{\y_f\y_f^H})   \\
    & = \mbox{Tr}(\Sigmahat_f + \yhat_f\yhat_f^H)\\
    & = \mbox{Tr}(\Sigmahat_f) + \yhat_f^H\yhat_f
  \\
      \Esp{\y \mid \vec{r} ; \vec{\theta}^{(m)}}{\y_f^H(\sigma_f^{2}\mat{W}^H\mat{W})\y}
      & =
      \Esp{\y \mid \vec{r} ;\vec{\theta}^{(m)}}{\mbox{Tr}((\sigma_f^{2}\mat{W}^H\mat{W})\y_f\y_f^H)}
      \\&
      = \mbox{Tr}((\sigma_f^2\mat{W}^H\mat{W})\Esp{\y \mid \vec{r} ;\vec{\theta}^{(m)}}{\y_f\y_f^H})   \\
      & = \mbox{Tr}((\sigma_f^2\mat{W}^H\mat{W})\Sigmahat_f + (\sigma_f^2\mat{W}^H\mat{W})\yhat_f\yhat_f^H)\\
      & = \mbox{Tr}((\sigma_f^2\mat{W}^H\mat{W})\Sigmahat_f) + \yhat_f^H(\sigma_f^2\mat{W}^H\mat{W})\yhat_f
  \end{aligned}
\end{equation}
Finally $Q_1$ reads,
\begin{equation}
  \begin{aligned}
  % \Q &=
  %   \Esp{\y \mid \x;\vec{\theta}^{(m)}}{\log \big(p\left(\x,\y ; \vec{\theta}\right) \big)}
  % \\
  % \Q &=
  %   \Esp{\y \mid \vec{r};\vec{\theta}^{(m)}}{\log(p(\vec{r} \mid \y ; \vec{\theta})) + \log(p(\y;\vec{\theta}))}
  % \\
  % \Q &=
  % C_1 + C_2 - \log\mid \sigma^2 \mat{I}_{4FN_{\mathcal{B}}}\mid -\ \log\mid \mat{I}_{\rho}\mid
  % \\ &-
  %   \frac{1}{\sigma^2}\Esp{\y \mid \vec{r} ;\vec{\theta}^{(m)}}{(\vec{r}-\vec{v}(\Z) -\mat{\Psi}\y)^H(\vec{r}-\vec{v}(\Z) -\mat{\Psi}\y)} -  \Esp{\y \mid \vec{r} ;\vec{\theta}^{(m)}}{\y^H\y}
  % \\
  % \Q &=
  % C - \log\mid \sigma^2 \mat{I}_{4FN_{\mathcal{B}}}\mid -\ \log\mid \mat{I}_{\rho}\mid - \frac{1}{\sigma^2}(\vec{r}-\vec{v})^H(\vec{r}-\vec{v})
  % \\ &+
  % \frac{2}{\sigma^2}\Re((\vec{r}-\vec{v})^H\mat{\Psi}\Esp{\y \mid \vec{r};\vec{\theta}^{(m)}}{\y})
  % - \frac{1}{\sigma^2} \Esp{\y \mid \vec{r};\vec{\theta}^{(m)}}{\y^H((\mat{\Psi}^H\mat{\Psi}))\y}
  % - \Esp{\y \mid \vec{r} ;\vec{\theta}^{(m)}}{\y^H\y}
  % \\
  \Q &=
  \rev{A} - F \log\mid \sigma^2 \mat{I}_{4B}\mid +
  \sum_{f=1}^F \frac{2}{\sigma^2}\Re((\vec{r}_f-\vec{v}_f)^H \sigma_f\mat{\W}\yhat_f) 
  \\ &
  - \frac{1}{\sigma^2}(\vec{r}_f-\vec{v}_f)^H(\vec{r}_f-\vec{v}_f)
  - \frac{\sigma_f^2}{\sigma^2}(\mbox{Tr}(\mat{W}^H\mat{W}\Sigmahat_f) + \yhat_f^H(\sigma_f^2\mat{W}^H\mat{W})\yhat_f)
  \\ &-
  \mbox{Tr}(\Sigmahat_f) - \yhat_f^H\yhat_f + 2\mbox{vec}(\mat{I})^H\yhat_f
  - \mbox{vec}(\mat{I})^H\mbox{vec}(\mat{I})
  \\ &=
  \rev{A} - F \log\mid \sigma^2 \mat{I}_{4B}\mid 
  - \sum_f \bigg( \frac{1}{\sigma^2}\norm{\vec{r}_f-\vec{v}_f - \sigma_f\mat{W}\yhat_f}_2^2
  \\ &+
  \norm{\yhat_f - \mbox{vec}(\mat{I})}_2^2
  + \frac{\sigma_f^2}{\sigma^2}\mbox{Tr}(\mat{W}^H\mat{W}^H\Sigmahat_f) + \mbox{Tr}(\Sigmahat_f) \bigg)
\end{aligned}
\end{equation}

\section{Maximization of $\Q$}
We compute the derivative of $Q_1(\vec{\theta} \mid \vec{\theta}^{(m)})$ w.r.t $\sigma_f$, for $f\in[1;F]$ :

  \begin{equation}
  \begin{aligned}
  \frac{\partial Q_1(\vec{\theta} \mid \vec{\theta}^{(m)})}{\partial \sigma_f}
  &=
  -\frac{2\sigma_f}{\sigma^2}\mbox{Tr}\big(\W^H\W (\Sigmahat_f + \yhat_f^H\yhat_f) \big) + \frac{2}{\sigma^2}\Re\bigg((\vec{r}_f-\v_f(\Z))^H\W\yhat_f\bigg)
  \end{aligned}
  \end{equation}
  We solve for $\cfrac{\partial Q_1(\vec{\theta} \mid \vec{\theta}^{(m)})}{\partial \sigma_f} = 0$ and we obtain 
  
  \begin{equation}
  \begin{aligned}
  \hat{\sigma}_f = \frac{\Re\bigg((\vec{r}_f-\v_f)^H\W\yhat_f\bigg)}{\mbox{Tr}\big(\W^H\W(\Sigmahat_f + \yhat_f\yhat_f^H) \big)}
  \end{aligned}
  \end{equation}
  $Q_1(\vec{\theta} \mid \vec{\theta}^{(m)})$ can be written in function of $\W$, 
  \begin{equation}
      \begin{aligned}
      Q_1(\vec{\theta} \mid \vec{\theta}^{(m)})
      &=
      A  + \frac{1}{\sigma^2} \sum_{f=1}^F  \bigg(2\Re\big((\vec{r}_f-\v_f)^H\sigma_f\W\yhat_f \big) - \mbox{Tr}(\sigma_f^2\W^H\W\Sigmahat_f) - \yhat_f^H\sigma_f^2\W^H\W\yhat_f \bigg)
      \\ &= A  + \frac{1}{\sigma^2} \sum_{f=1}^F  \bigg(
      \mbox{Tr}\left(\W\sigma_f\yhat_f(\vec{r}_f-\v_f)^H\right) +
      \mbox{Tr}\left(\W^H\sigma_f\yhat_f^H(\vec{r}_f-\v_f)\right)
      \\ &-
       \mbox{Tr}\left(\sigma_f^2\W^H\W(\Sigmahat_f +\yhat_f\yhat_f^H) \right) \bigg)
      \end{aligned}
  \end{equation}
  The derivatives of $Q_1$ w.r.t $\W$ is computed using Wirtinger derivative \cite{hjorougnes07}
  
  \begin{equation}
    \begin{aligned}
    \frac{\partial}{\partial\W} \bigg( \mbox{Tr}\left(\W\sigma_f\yhat_f(\vec{r}_f-\v_f)^H\right) \bigg) &= \left(\sigma_f\yhat_f(\vec{r}_f-\v_f)^H \right)^T
  \\
    \frac{\partial}{\partial\W} \bigg( \mbox{Tr}\left(\sigma_f\yhat_f(\vec{r}_f-\mat{v_f})^H \W^H \right) \bigg) &= \mat{0}_{4B\times r}
  \\
    \frac{\partial}{\partial\W} \bigg(\mbox{Tr}\left(\sigma_f^2\W^H\W(\Sigmahat_f +\yhat_f\yhat_f^H) \right) \bigg) &= \sigma_f^2 \W^* (\Sigmahat_f +\yhat_f\yhat_f^H)^T.
    \end{aligned}
  \end{equation}
  Consequently,
  
\begin{equation}
  \begin{aligned}
  \frac{\partial}{\partial\W}(Q_1(\vec{\theta} \mid \vec{\theta}^{(m)})) &= \sum_{f=1}^F \bigg(\left(\sigma_f\yhat_f(\vec{r}_f-\v_f)^H \right)^T -  \sigma_f^2 \W^* (\Sigmahat_f +\yhat_f\yhat_f^H)^T\bigg)
\\
 &= \sum_{f=1}^F \sigma_f \left(\yhat_f(\vec{r}_f-\v_f)^H \right)^T
  - \W^* \sum_{f=1}^F \sigma_f^2 \left(\Sigmahat_f +\yhat_f\yhat_f^H \right)^T
\end{aligned}
\end{equation}
  Finally, solving for $\cfrac{\partial Q_1(\vec{\theta} \mid \vec{\theta}^{(m)})}{\partial\W} = 0$ leads to,
  \begin{equation}
    \begin{aligned}
  \W^* &=  \bigg(\sum_{f=1}^F \sigma_f \left(\yhat_f(\vec{r}_f-\v_f)^H \right)^T \bigg) \bigg(\sum_{f=1}^F \sigma_f^2(\Sigmahat_f +\yhat_f\yhat_f^H)^T \bigg)^{-1}
\\
  \W &=  \bigg(\sum_{f=1}^F \sigma_f \left(\yhat_f(\vec{r}_f-\mat{v_f})^H \right)^H \bigg) \bigg(\sum_{f=1}^F \sigma_f^2 \left(\Sigmahat_f +\yhat_f\yhat_f^H\right)^H \bigg)^{-1}
\end{aligned}
\end{equation}

\section{Proof of proposition 2}

From (\ref{complete-set-2}), 
$$
\u_i \sim CN(\v_i(\Z), \beta_i \sigma^2 \mat{I}_{4FN_{\mathcal{B}}}) \text{ with } \sum_{i=1}^D \beta_i = 1
$$
% $$ \text{ where }\v_i(\Z)= \{(\mat{B}^*(f) \otimes \mat{B}_f)(\Z^*_{i,q} \otimes \Z_{i,p} )\vec{c}_{i,f}\}_{f,p,q}. $$
The log-likelihood of the complete data can be written 
\begin{equation}
  \log(p(\u_i ; \vec{\theta})) =
  C_{u_{i}} - \log\mid \beta_i\sigma^2 \mat{I}_{4FN_{\mathcal{B}}} \mid - \frac{1}{\beta_i\sigma^2} (\u_i - \v_i (\Z))^H (\u_i - \v_i(\Z)).
\end{equation}
Thus, $Q_2$ can be expressed as follows,
\begin{equation}
  \begin{aligned}
  Q_2(\vec{\theta} \mid \vec{\theta}^{(m)})
  &=
  \sum_{i=1}^D
  \Esp{\u_i \mid \vec{r};\vec{\theta}^{(m)}}{\log(p(\u_i ; \vec{\theta}))}
  \\ &=
  \sum_{i=1}^D
  \Esp{\u_i \mid \vec{r}}{
          C_{u_{i}} - \log\mid \beta_i\sigma^2 \mat{I}_{4FN_{\mathcal{B}}} \mid - \frac{1}{\beta_i\sigma^2} (\u_i - \v_i(\Z))^H (\u_i - \v_i(\Z))}
\\ &=
  \sum_{i=1}^D
    C_{u_{i}} - \log\mid \beta_i\sigma^2 \mat{I}_{4FN_{\mathcal{B}}} \mid
  -
   \frac{1}{\beta_i \sigma^2}\Esp{\u_i \mid \vec{r}}{(\u_i - \vec{v}_i(\Z))^H(\rev{\vec{r}_i} - \vec{v}_i(\Z))}
\\ &=
  \sum_{i=1}^D
  C_{u_{i}} - \log\mid \beta_i\sigma^2 \mat{I}_{4FN_{\mathcal{B}}} \mid
-
  \frac{1}{\beta_i \sigma^2}\Esp{\u_i \mid \vec{r}}{\u_i^H\u_i}
+
  \frac{2}{\beta_i \sigma^2}\Re\bigg(\v_i(\Z)^H \Esp{\u_i \mid \vec{r}} {\u_i}\bigg)
\\ &-
  \frac{1}{\beta_i \sigma^2}\v_i^H\v_i.
  \end{aligned}
\end{equation}

\def\Sigmahatui{{\hat{\mat{\Sigma}}_{\u_i}}}
\def\uihat{{ \hat{\u}_i }}
\noindent Let us note $\uihat = \Esp{ \u_i \mid \vec{r};\vec{\theta}^{(m)}}{\u_i}$
and $\Sigmahatui = \Esp{ \u_i \mid \vec{r};\vec{\theta}^{(m)}}{(\u_i-\uihat)(\u_i- \uihat)^H}$ the mean and covariance
of the complete data given the oberved data and the parameter vector.
We consider \rev{$\vec{z}_i = [\u_i, \vec{r}_i]$} being a multivariate Gaussian law. The mean of  $\u_i$ and $\vec{r}$ are respectively
$$
\vec{\mu_{\u_i}}= \vec{v}_i(\Z) \text{, } \vec{\mu_{\vec{r}}} = \vec{v}(\mat{Z}) + \mat{\Psi}(\mbox{vec}(\eye{}) \otimes \mathbf{1}_F).
$$

\begin{equation}
  \begin{aligned}
    \mat{\Sigma}_{\u_i \u_i}
    &=
     \Esp{\u_i}{(\u_i - \vec{\mu}_{\u_i})(\u_i - \vec{\mu}_{\u_i})^H}
    =
     \Esp{\u_i}{(\u_i - \vec{v}_i(Z))(\u_i - \vec{v}_i(Z))^H}
    =
     \beta_i \sigma^2 \mat{I}_{4FN_{\mathcal{B}}}
\\
    \mat{\Sigma}_{\vec{r} \vec{r}}
    &=
    \Esp{\vec{r}}{(\vec{r} - \vec{\mu_{\vec{r}}})(\vec{r} - \vec{\mu_{\vec{r}}})^H}
    =
    \Esp{\vec{r}}{(\vec{r}-\vec{v}(\mat{Z}))(\vec{r}-\vec{v}(\mat{Z}))^H}
    =
      (\sigma^2 \mat{I} + \mat{\Psi}\mat{\Psi}^H)
\\
    \mat{\Sigma_{\vec{r} \u_i}}
    &=
    \mat{\Sigma_{\u_i \vec{r}}}^H
    =
     \Esp{}{(\u_i - \vec{\mu}_{\u_i})(\vec{r} - \vec{\mu}_{\vec{r}})^H}\\
    &=
    \Esp{}{(\u_i - \v_i(\Z))(\sum_{k=1}^D \left(\u_k - \v_k(\Z) \right) + \mat{\Psi}\vec{y})^H} \\
    &=
    \sum_{k=1}^D
    \Esp{}{(\u_i - \v_i(\Z))(\vec{r}_k - \vec{v}_k)^H}\\
    \end{aligned}
\end{equation}
Since the $\u_i$ are considered independant from one another
\begin{equation}
    \begin{aligned}
      \mat{\Sigma}_{\vec{r} \u_i}
      =
      \mat{\Sigma}_{\u_i \vec{r}}^H
      &=
       \mat{\Sigma}_{\u_i \u_i}
      +
       \sum_{k=1, k\neq i}^D
       \Esp{}{(\u_i - \vec{v}_i)}
       \Esp{}{(\u_k - \vec{v}_k)} \\
    &= \mat{\Sigma}_{\u_i \u_i}\\
    \end{aligned}
\end{equation}

\noindent Using \cite[p36]{Anderson03} we can compute the analytical expression for $\Sigmahatui$ and $\uihat$ as

\begin{equation}
  \begin{aligned}
    \uihat
    &=
    \vec{\mu}_{\u_i} + \mat{\Sigma}_{\u_i \vec{r}}\mat{\Sigma}_{\vec{r} \vec{r}}^{-1}(\vec{r} - \mu_{\vec{r}})
    =
     \vec{v}_i(\Z) + \beta_i \sigma^2 (\sigma^2 \mat{I} + \mat{\Psi}\mat{\Psi}^H)^{-1}(\vec{r} - \vec{v}(Z) - \mat{\Psi}(\mbox{vec}(\eye{}) \otimes \mathbf{1}_F))
\\
    \Sigmahatui
    &=
     \mat{\Sigma}_{\u_i \u_i} + \mat{\Sigma}_{\u_i \vec{r}}\mat{\Sigma}_{\vec{r}\vec{r}}^{-1}\mat{\Sigma}_{\vec{r} \u_i}
    =
      \beta_i\sigma^2 \mat{I}_{4FN_{\mathcal{B}}} - \beta_i^2 \sigma^4 (\sigma^2 \mat{I} + \mat{\Psi}\mat{\Psi}^H)^{-1}
  \end{aligned}
\end{equation}
We can then express 
\begin{equation}
\begin{aligned}
    \Esp{\u_i \mid \vec{r}}{\u_i}
    &=
     \uihat
     \\
    \Esp{\u_i \mid \vec{r}}{\u_i^H\u_i}
    &=
     \mbox{Tr}\left(\Sigmahatui \right) + \uihat^{H}\uihat.
\end{aligned}
\end{equation}
Finally, 
\begin{equation}
\begin{aligned}
  Q_2(\vec{\theta} \mid \vec{\theta}^{(m)})
  &=
    \sum_{i=1}^D C_{u_{i}} -\log\mid \beta_i\sigma^2 \mat{I}_{4FN_{\mathcal{B}}} \mid - \frac{1}{\beta_i \sigma^2} \norm{\uihat - \v_i(\Z)}_F^2 - \frac{1}{\beta_i \sigma^2} \mbox{Tr}(\Sigmahatui).
\end{aligned}
\end{equation}

\section{Maximization of $Q_2(\vec{\theta}, \vec{\theta^{(m)}})$}

We compute the derivatives of $Q_2$ w.r.t $\sigma^2$:

\begin{equation}\label{appendix:dQ2dsig2}
  \begin{aligned}
    \frac{\partial{Q_2}}{\sigma^2} &= \sum_{i=1}^D -\frac{4FN_{\mathcal{B}}}{\sigma^2} + \frac{1}{\beta_i\sigma^4}\left( \norm{\uihat - \v_i(\Z)}_F^2 + \mbox{Tr}(\Sigmahatui) \right)
    \\
    &= \frac{1}{\sigma^4}\sum_{i=1}^D -4FN_{\mathcal{B}}\sigma^2 + \frac{1}{\beta_i}\left( \norm{\uihat - \v_i(\Z)}_F^2 + \mbox{Tr}(\Sigmahatui) \right)
    \\
    & = \frac{1}{\sigma^4}\left( -4FN_{\mathcal{B}}D\sigma^2  + \sum_{i=1}^D \frac{1}{\beta_i}\left( \norm{\uihat - \v_i(\Z)}_F^2 + \mbox{Tr}(\Sigmahatui) \right) \right)
  \end{aligned}
\end{equation}
The update for $\sigma^2$ is then derived from \ref{appendix:dQ2dsig2},

\begin{equation}
  \begin{aligned}
    \sigma^{2} = \frac{1}{4FN_{\mathcal{B}}D}  \sum_{i=1}^D \frac{1}{\beta_i}\left( \norm{\uihat - \v_i(\Z)}_F^2 + \mbox{Tr}(\Sigmahatui) \right)
  \end{aligned}
\end{equation}
In order to derive the update for $\Z$, $Q_2$ is express to highlight the dependence with $\Z$,

  \begin{equation}
  \begin{aligned}
    Q_2(\Z)
    &=
      A -
      \sum_{i=1}^D \frac{1}{\beta_i \sigma^2} \norm{\uihat - \v_i(\Z)}_F^2,
  \end{aligned}
  \end{equation}
  where $A$ is a constant independant of $\mat{Z}$.
  Since $\uihat$ and $\v_{i}(\Z_i)$ can both be divided into $F$ sub-vectors, they can be written as
  $\uihat = [\vec{\hat{u}}_{i,1}^T, ..., \vec{\hat{u}}_{i,F}^T]^T$ and 
  $\v_{i}(\Z) = [\v_{i,1}(\Z)^T, ..., \v_{i,F}(\Z)^T]^T$.
  Thus $Q_2$ can be expressed as follows,
  $$
  Q_2(\vec{\theta} \mid \vec{\theta}^{(m)})
  =
   A
   -
    \sum_{i=1}^D \sum_{f=1}^F \frac{1}{\beta_i \sigma^2} \norm{\vec{\hat{u}}_{if} - \v_{i,f}(\Z)}_F^2.
  $$
  We consider
  $$
  \phi(\Z_i) =
    \sum_{f=1}^F
      \frac{1}{\beta_i \sigma^2} \norm{\vec{\hat{u}}_{if} - \v_{i,f}(\Z)}_F^2
  $$
  where $\v_{i,f}$ is the $i^{th}$ source noise free component. Although, $\v_{i,f}$ can be
  partitioned in $N_{\mathcal{B}}=\frac{N(N-1)}{2}$ 4-dimension vectors
  $\v_{i,f} = [\v_{i,f,12}^T, \v_{i,f,13}^T, ..., \v_{i,f,N(N-1)}^T]^T$  and
  $\v_{i,f,pq} = \left(\mat{B}_f^* \otimes \mat{B}_f  \right) (\Z_{i,q}^* \otimes \Z_{i,p})\mbox{vec}(\mat{C}_{i,f})$.
  We note as well $\hat{\u}_{i,f,pq}$ the $4\times 1$ vector such that
  $
  \hat{\vec{u}}_{i,f} = [\hat{\vec{u}}_{i,f,12}, \hat{\vec{u}}_{i,f,13} ... ,
  \hat{\vec{u}}_{i,f,(N-1)N}].
  $  
  Thus $\phi(\Z_i)$ can now be written 

  $$
  \phi(\Z_i) = \sum_{f=1}^F \sum_{(p,q)\in[1;N]^2 \mid p<q}
  \norm{\vec{\hat{u}}_{i,f,pq} - \v_{i,f,pq}(\Z_i)}_F^2
  $$
  where $\Z_i$ is a $2KN \times 2$ matrix that can be expressed using $2K \times 2$ matrix
  
  $$
  \Z_{i} =
  \begin{bmatrix}
  \Z_{i,1} \\
  \Z_{i,2} \\
  ...       \\
  \Z_{i,N}
  \end{bmatrix}
  \qquad \text{where }
   \Z_{i,p}\in\mathcal{M}_{2K,2}
  \mbox{ and }
  \Z_{i,p} =
  \begin{bmatrix}
  \Z_{1,i,p} \\
  \Z_{2,i,p} \\
  ...       \\
  \Z_{K,i,p}
  \end{bmatrix}
  \qquad \text{where }
  \Z_{k,i,p}\in\mathcal{M}_{2,2}
  $$
  Let us recall that
  
  \begin{equation}
    \begin{aligned}
      \v_{i,f,pq}
       &=
        \big((\mat{B}_f^* \Z_{i,q}^*) \otimes (\mat{B}_f\Z_{i,p})\big)\mbox{vec}(\mat{C}_{i,f})
      \\&=
        \mbox{vec}( \mat{B}_f\Z_{i,p} \mat{C}_{i,f} \Z_{i,q}^H\mat{B}_f^H)
    \end{aligned}
  \end{equation}
  By considering $\hat{\mat{U}}_{i,f,pq} = \mbox{unvec}(\vec{\hat{u}}_{i,f,pq})$, we can express $\phi(\Z_i)$ as
  
  \begin{equation}
    \begin{aligned}
      \phi(\Z_i)
      =
      \sum_{f=1}^F \sum_{(p,q)\in[1;N]^2 \mid p<q}
      \norm{\hat{\mat{U}}_{i,f,pq} -  \mat{B}_f\Z_{i,p} \mat{C}_{i,f} \Z_{i,q}^H\mat{B}_f^H}_F^2.
    \end{aligned}
  \end{equation}
  $\phi(\Z_i)$ is minimized w.r.t to $\Z_i$ with a block coordinate descent \cite{FHHT07} by minimizing
  recursevely $\phi(\Z_i)$ w.r.t $\Z_{i,p}$ for a given $\Z_{i,q}$ with $p \neq q$.
  Let $p\in[1:N]$, we note
  
  $$
  f_{i,f,pq}(\Z_{i,p})
  = \norm{\hat{\mat{U}}_{i,f,pq} -  \mat{B}_f\Z_{i,p} \mat{C}_{i,f} \Z_{i,q}^H\mat{B}_f^H}_F^2
  $$
  
  $$
  g_{i,f,pq}(\Z_{i,p})
  = \norm{\hat{\mat{U}}_{i,f,qp} -  \mat{B}_f\Z_{i,q} \mat{C}_{i,f} \Z_{i,p}^H\mat{B}_f^H}_F^2.
  $$
  The cost function $\phi(\Z_{i,p})$ becomes, 
  \begin{equation}
    \begin{aligned}
      \phi(\Z_{i,p})
      =
      A
      +
      \sum_{f=1}^F \sum_{q=p+1}^{N} f_{i,\rev{f},pq}(\Z_{i,p})
      +
      \sum_{f=1}^F \sum_{q=1}^{p-1} g_{i,\rev{f},pq}(\Z_{i,p}).
    \end{aligned}
  \end{equation}
  The partial derivatives of $f$ and $g$ w.r.t $\Z_{i,p}$ are computed using \cite{hjorougnes07}
  $$
      \frac{\partial f_{i,f,pq}}{\Z_{i,p}}
      =
      \mat{B}_f^T
      (\hat{\mat{U}}_{i,f,pq}^*  -
      \mat{B}_f^*\Z_{i,p}^*\mat{F}_{i,f,q}^H)\mat{F}_{i,f,q}
  $$

  $$
      \frac{\partial g_{i,f,pq}}{\Z_{i,p}}
      =
      \mat{B}_f^T
      (\hat{\mat{U}}_{i,f,qp}^T  -
      \mat{B}_f^* \Z_{i,p}^* \mat{G}_{i,f,q}^H)\mat{G}_{i,f,q}.
  $$
  Thus, the partial derivative of $\phi(\Z_{i,p})$ w.r.t $\Z_{i,p}$ can be expressed,
  
  \begin{equation}
    \begin{aligned}
      \frac{\partial\phi}{\Z_{i,p}}
    &=
      \sum_{f=1}^F \sum_{q=p+1}^{N}
      \mat{B}_f^T
      (\hat{\mat{U}}_{i,f,pq}^*  -
      \mat{B}_f^*\Z_{i,p}^*\mat{F}_{i,f,q}^H)\mat{F}_{i,f,q}
      \\ &+
      \sum_{f=1}^F \sum_{q=1}^{p-1}
      \mat{B}_f^T
      (\hat{\mat{U}}_{i,f,qp}^T  -
      \mat{B}_f^* \Z_{i,p}^* \mat{G}_{i,f,q}^H)\mat{G}_{i,f,q}
    \\ &=
      \sum_{f=1}^F
      \sum_{q=p+1}^{N}\mat{B}_f^T \hat{\mat{U}}_{i,f,pq}^*\mat{F}_{i,f,q}
      +
      \sum_{f=1}^F
      \sum_{q=1}^{p-1} \mat{B}_f^T \hat{\mat{U}}_{i,f,qp}^T\mat{G}_{i,f,q}
      \\&-
      \sum_{f=1}^F
      \sum_{q=p+1}^{N}
      \mat{B}_f^T \mat{B}_f^*\Z_{i,p}^*\mat{F}_{i,f,q}^H \mat{F}_{i,f,q}
      \\&-
      \sum_{f=1}^F
      \sum_{q=1}^{p-1}
      \mat{B}_f^T \mat{B}_f^* \Z_{i,p}^* \mat{G}_{i,f,q}^H \mat{G}_{i,f,q}.
    \end{aligned}
  \end{equation}
  Let us note
  $$
  \mat{T}_{i,p} = \sum_{f=1}^F \sum_{q=p+1}^{N} \mat{B}_f^T \hat{\mat{U}}_{i,f,pq}^*\mat{F}_{i,f,q}
  +
   \sum_{f=1}^F \sum_{q=1}^{p-1} \mat{B}_f^T \hat{\mat{U}}_{i,f,qp}^T\mat{G}_{i,f,q}.
  $$
  The partial derivative of $\phi(\Z_{i,p})$ w.r.t $\Z_{i,p}$ becomes
  
  \begin{equation}
    \begin{aligned}
      \frac{\partial\phi}{\Z_{i,p}}
    &=
        \mat{T}_{i,p}
        -
        \sum_{f=1}^F
        \sum_{q=p+1}^{N}
        \mat{B}_f^T \mat{B}_f^*\Z_{i,p}^*\mat{F}_{i,f,q}^H \mat{F}_{i,f,q}
        -
        \sum_{f=1}^F
        \sum_{q=1}^{p-1}
        \mat{B}_f^T \mat{B}_f^* \Z_{i,p}^* \mat{G}_{i,f,q}^H \mat{G}_{i,f,q}.
    \end{aligned}
  \end{equation}
  Setting the partial derivative of $\phi(\Z_{i,p})$ w.r.t $\Z_{i,p}$ to zero is equivalent to
  setting the partial derivative of $\mbox{vec}(\phi(\Z_{i,p}))$ w.r.t $\mbox{vec}(\Z_{i,p})$ to zero,
  
  \begin{equation}
    \begin{aligned}
      \mbox{vec}(\frac{\partial\phi}{\Z_{i,p}})
    &=
      \mbox{vec}(\mat{T}_{i,p})
      -
      \sum_{f=1}^F \sum_{q=p+1}^{N}
      \mbox{vec}(\mat{B}_f^T \mat{B}_f^*\Z_{i,p}^*\mat{F}_{i,f,q}^H \mat{F}_{i,f,q})
    \\ &-
      \sum_{f=1}^F \sum_{q=1}^{p-1}
      \mbox{vec}(\mat{B}_f^T \mat{B}_f^* \Z_{i,p}^* \mat{G}_{i,f,q}^H \mat{G}_{i,f,q})
  \\&=
      \mbox{vec}(\mat{T}_i)
      -
      \sum_{f=1}^F \sum_{q=p+1}^{N}
      \bigg(\mat{F}_{i,f,q}^T\mat{F}_{i,f,q}^* \otimes  \mat{B}_f^T \mat{B}_f^* \bigg) \mbox{vec}(\Z_{i,p}^*)
    \\ &-
      \sum_{f=1}^F \sum_{q=1}^{p-1}
      \bigg(\mat{G}_{i,f,q}^T\mat{G}_{i,f,q}^* \otimes  \mat{B}_f^T \mat{B}_f^* \bigg) \mbox{vec}(\Z_{i,p}^*).
    \end{aligned}
  \end{equation}
  Finally, we obtain a solution for $\Z_{i,p}$ such that $\mbox{vec}(\cfrac{\partial\phi}{\Z_{i,p}}) = 0$,
  
  $$
      \mbox{vec}(\Z_{i,p}^*) = (\mat{M}_{i,p} + \tilde{\mat{M}}_{i,p})^{-1}\mbox{vec}(\mat{T}_{i,p}),
  $$

  $$
  \text{with }
  \mat{M}_{i,p} = \sum_{f=1}^F \sum_{q=p+1}^{N}
  \bigg(\mat{F}_{i,f,q}^T\mat{F}_{i,f,q}^* \otimes  \mat{B}_f^T \mat{B}_f^* \bigg)
  $$
  
  $$
  \tilde{\mat{M}}_{i,p} = \sum_{f=1}^F \sum_{q=1}^{p-1}
  \bigg(\mat{G}_{i,f,q}^T\mat{G}_{i,f,q}^* \otimes  \mat{B}_f^T \mat{B}_f^* \bigg)
  $$
  
  $$
  \mat{T}_{i,p} = \sum_{f=1}^F \sum_{q=p+1}^{N} \mat{B}_f^T \hat{\mat{U}}_{i,f,pq}^*\mat{F}_{i,f,q}
  +
   \sum_{f=1}^F \sum_{q=1}^{p-1} \mat{B}_f^T \hat{\mat{U}}_{i,f,qp}^T\mat{G}_{i,f,q}
  $$

  $$
  \mat{F}_{i,f,q} = \mat{B}_f^* \Z_{i,q}^* \mat{C}_{i,f}^T
  $$
  
  $$
  \mat{G}_{i,f,q} = \mat{B}_f^* \Z_{i,q}^* \mat{C}_{i,f}^*.
  $$

\bibliographystyle{ieeetr} 
\bibliography{refs}

\end{document}